\def\version{April 21, 2016}

\documentclass[12pt]{article}

\title
{Finite-order correlation length for 4-dimensional \\
weakly self-avoiding walk and $|\varphi|^4$ spins}

\author{Roland Bauerschmidt\thanks{Department of Mathematics,
Harvard University, 1 Oxford Street, Cambridge, MA 02138, USA.
{\tt brt@math.harvard.edu}}, Gordon Slade\thanks{Department of Mathematics,
University of British Columbia, Vancouver, BC, Canada V6T 1Z2.
{\tt slade@math.ubc.ca}, {\tt atomberg@math.ubc.ca}, {\tt bwallace@math.ubc.ca}},
\\ Alexandre Tomberg$^\dagger$ and Benjamin C. Wallace$^\dagger$
}

\def\macrosPb{}
\def\macrosHarxiv{}
\usepackage{amsfonts}
\usepackage{amsmath,amssymb,amsthm}
\usepackage{appendix}
\usepackage{bbm} 
\usepackage{amsbsy}
\usepackage{enumerate}
\usepackage{cite}
\usepackage{enumerate}

\InputIfFileExists{./macros_local.tex}{}{}

\ifdefined\macrosPa
  \usepackage[textwidth=465pt,textheight=650pt,centering]{geometry} 
\else\ifdefined\macrosPb
  \usepackage[textwidth=500pt,textheight=650pt,centering]{geometry} 
\fi\fi

\ifdefined\macrosS


  \usepackage{mathptmx}
  \DeclareMathAlphabet{\mathcal}{OMS}{cmsy}{m}{n}
\fi

\ifdefined\macrosBirk
\else
\usepackage[dvips]{graphicx}
\fi

\ifdefined\macrosSB

\def\UseSection{
        \numberwithin{equation}{section}
	\theoremstyle{plain}
        \newtheorem{theorem}    {Theorem}[section]
        \DefineTheorems 
}

\def\DefineTheorems{
	
	\newtheorem{lemma}      [theorem] {Lemma}
	
	\newtheorem{prop}       [theorem] {Proposition}
	
	\newtheorem{cor}        [theorem] {Corollary}

	\theoremstyle{definition}
	\newtheorem{defn}       [theorem] {Definition}

	\newtheorem{rk} 	[theorem] {Remark}
	\theoremstyle{definition}

}

\newcommand{\bt}   {\begin{theorem}}
\newcommand{\et}   {\end  {theorem}}
\newcommand{\bl}   {\begin{lemma}}
\newcommand{\el}   {\end  {lemma}}
\newcommand{\bp}   {\begin{prop}}
\newcommand{\ep}   {\end  {prop}}
\newcommand{\bc}   {\begin{cor}}
\newcommand{\ec}   {\end  {cor}}
\newcommand{\bd}   {\begin{defn}}
\newcommand{\ed}   {\end  {defn}}

\newcommand{\ba}   {\begin{array}}
\newcommand{\ea}   {\end  {array}}
\newcommand{\be}   {\begin{enumerate}}
\newcommand{\ee}   {\end  {enumerate}}
\newcommand{\bi}   {\begin{itemize}}
\newcommand{\ei}   {\end  {itemize}}

\def\eq#1\en{\begin{equation}#1\end{equation}}  
\def\eqsplit#1\ensplit{
	\begin{equation}\begin{split}#1\end{split}\end{equation}
	}
\def\eqalign#1\enalign{
	\begin{align}#1\end{align}
	}
\def\eqmul#1\enmul{
	\begin{multline}#1\end{multline}
	}
\newcommand{\eqarrstar} {\begin{eqnarray*}} 
\newcommand{\enarrstar} {\end{eqnarray*}} 
\newcommand{\eqarray}   {\begin{eqnarray}} 
\newcommand{\enarray}   {\end{eqnarray}} 
\newcommand{\nnb}	{\nonumber \\} 

\newcommand{\lbeq}[1]  {\label{e:#1}}
\newcommand{\refeq}[1] {\eqref{e:#1}}    

\makeatletter
\newcommand{\labelcounter}[2]{{%
	\stepcounter{#1}
	\protected@write\@auxout{}%
	{\string\newlabel{#2}{{\csname the#1\endcsname}{\thepage}}}%
	{\ref{#2}}
	}}
\makeatother
\newcommand{\Ebold} {{\mathbb E}}

\newcommand{\Rbold} {{\mathbb R}}

\newcommand{\Zbold} {{\mathbb Z}}

\newcommand{\Bcal}   {\mathcal{B}} 
 
\newcommand{\Dcal}   {\mathcal{D}} 
 
\newcommand{\Fcal}   {\mathcal{F}} 
\newcommand{\Gcal}   {\mathcal{G}}

\newcommand{\Kcal}   {\mathcal{K}}

\newcommand{\Ncal}   {\mathcal{N}} 
 
\newcommand{\Pcal}   {\mathcal{P}}

\newcommand{\Scal}   {\mathcal{S}}

\newcommand{\Vcal}   {\mathcal{V}} 
\newcommand{\Wcal}   {\mathcal{W}} 

\newcommand{\Ycal}   {\mathcal{Y}}

\newcommand{\Rd}    {{ {\Rbold}^d}}
\newcommand{\Zd}    {{ {\Zbold}^d }}

\newcommand{\spose}[1] {{\hbox to 0pt{#1\hss}} }
\newcommand{\ltapprox} {\mathrel{\spose{\lower 3pt\hbox{$\mathchar"218$}}
 \raise 2.0pt\hbox{$\mathchar"13C$}}}
\newcommand{\gtapprox} {\mathrel{\spose{\lower 3pt\hbox{$\mathchar"218$}}
 \raise 2.0pt\hbox{$\mathchar"13E$}}}

\else
\fi

\UseSection   
\usepackage[usenames]{color}

\definecolor{bw}{RGB}{240, 120, 0}
\definecolor{at}{rgb}{0.0, 0.5, 0.0} 

\newcommand{\Tay}{{\rm Tay}}

\newcommand{\LT}{{\rm Loc}  }

\newcommand{\DV}{\Dcal}
\newcommand{\DVa}{\alpha}

\renewcommand{\to} {\rightarrow}

\newcommand{\R}{\Rbold}
\newcommand{\Z}{\Zbold}

\newcommand{\C}{\mathbb{C}}

\newcommand{\1}{\mathbbm{1}}

\newcommand{\Ex}{\mathbb{E}}

\newcommand{\Econst}{\alpha_{\Ebold}}

\newcommand{\chicCov}{{\chi}}
\newcommand{\ellconst}{\mathfrak{c}}

\newcommand{\Gtilp}{\gamma}

\newcommand{\pair}[1]{\langle #1 \rangle}

\newcommand{\pt}{{\rm pt}}

\newcommand{\Ipttil}{\tilde{I}_{\rm pt}}

\newcommand{\Vpt}{V_{\rm pt}}

\newcommand{\h}{\mathfrak{h}}
\ifdefined\macrosSB \else

\fi

\newcommand{\gbar}{\bar{g}}

\newcommand{\ggen}{\tilde{g}}

\newcommand{\chigen}{\tilde{\chi}}

\newcommand{\domRG}{\mathbb{D}}

\newcommand{\pp}{a}
\newcommand{\qq}{b}
\newcommand{\sigmaa}{\sigma}
\newcommand{\sigmab}{\bar{\sigma}}

\newcommand{\half}{\textstyle{\frac 12}}

\newcommand{\epdV}{\bar{\epsilon}}

\newcommand{\phifour}{$|\varphi|^4$}

\ifdefined\macrosH
  \usepackage{xr-hyper}
  \usepackage{hyperref}
  \hypersetup{hypertexnames=false}
  \hypersetup{colorlinks,citecolor=blue,linkcolor=blue}  

\else\ifdefined\macrosHarxiv
  \usepackage{xr-hyper}
  \usepackage{hyperref}
  \hypersetup{hypertexnames=false}

\else
  \newcommand{\texorpdfstring}[2]{#1}
  \usepackage{xr}
\fi\fi

\renewcommand{\chicCov}{\vartheta}
\renewcommand{\chigen}{\tilde{\vartheta}}
\newcommand{\clo}{p}

\newcommand{\old}{\mathrm{old}}

\date\version

\begin{document}

\maketitle

\begin{abstract}
We study the 4-dimensional $n$-component $|\varphi|^4$ spin model for all integers $n \ge 1$,
and the 4-dimensional continuous-time weakly self-avoiding walk which corresponds
exactly to the case $n=0$ interpreted as a supersymmetric spin model.
For these models, we analyse
the correlation length of order $p$,
and prove the existence of a logarithmic correction to mean-field scaling,
with power $\frac 12\frac{n+2}{n+8}$, for all $n \ge 0$ and $p>0$.
The proof is based on an improvement of a rigorous renormalisation group method
developed previously.
\end{abstract}


\section{Introduction and main results}

\subsection{Introduction}

Recently, using a rigorous renormalisation group method
\cite{BS-rg-norm,BS-rg-loc,BBS-rg-pt,BS-rg-IE,BS-rg-step,BBS-rg-flow},
the critical behaviour of the 4-dimensional
$n$-component $|\varphi|^4$ spin model \cite{BBS-phi4-log,
ST-phi4} and the 4-dimensional
continuous-time weakly self-avoiding walk \cite{BBS-saw4-log,BBS-saw4,ST-phi4} has been
analysed.
The latter model corresponds to the case $n=0$ via an exact identity which represents
the weakly self-avoiding walk as a supersymmetric field theory with quartic self-interaction.
A typical result in this work is that for all $n \ge 0$ the susceptibility diverges as
$\varepsilon^{-1} (\log \varepsilon^{-1})^{\frac{n+2}{n+8}}$, in the limit $\varepsilon
\downarrow 0$ describing approach to the critical point.
Related results have been obtained for the pressure, the specific heat,
the critical two-point function, and other quantities.
The existence of such logarithmic corrections to scaling for dimension~4 was predicted
about 45 years ago in the physics literature \cite{LK69,BGZ73,WR73}.
For $n=1$, the existence of logarithmic corrections was proven rigorously
about 30 years ago in \cite{Hara87,HT87}.

A missing aspect
in the analysis of critical scaling in
\cite{BBS-phi4-log,BBS-saw4-log,BBS-saw4,ST-phi4} is a determination of
the divergence
of correlation length scales as the critical point is approached.
A natural measure of length scale is the correlation length $\xi$ defined as the reciprocal
of the exponential decay rate of the two-point function.  We do not study this correlation
length (which was however studied in \cite{HT87} for the case $n=1$).
Instead, we study $\xi_p$, the \emph{correlation length of order $p$}, for all $p>0$,
and prove that its divergence takes the form
$\varepsilon^{-\frac{1}{2}} (\log \varepsilon^{-1})^{\frac{1}{2}\frac{n+2}{n+8}}$.
The independence of $p$ in the exponents
exemplifies the conventional wisdom that in critical phenomena
all naturally defined length scales should exhibit the same
asymptotic behaviour.
The correlation length $\xi$ is predicted to diverge in the same manner, but
our method would require further development to prove this.

\subsection{Definitions of the models}

Before defining the models, we establish some notation.
Let $L > 1$ be an integer (which we will need to fix large).
Consider the sequence $\Lambda=\Lambda_N = \Z^d/(L^N\Z^d)$ of
discrete $d$-dimensional tori of side lengths $L^N$,
with $N \to \infty$ corresponding to the infinite volume limit $\Lambda_N \uparrow \Z^d$.
Throughout the paper, we only consider $d=4$, but we sometimes write $d$ instead of $4$
to emphasise the role of dimension.
For any of the $2d$ unit vectors $e \in \Z^d$, we define
the discrete gradient of a function $f:\Lambda_N \to \R$
by $\nabla^e f_x = f_{x + e} - f_x$, and
the discrete Laplacian by
$\Delta = -\frac{1}{2}\sum_{e\in\Z^d:|e| = 1}\nabla^{-e} \nabla^{e}$.
The gradient and Laplacian operators act component-wise on vector-valued functions.
We also use the discrete Laplacian $\Delta_{\Zd}$ on $\Zd$, and the continuous Laplacian
$\Delta_{\Rd}$ on $\Rd$.

\subsubsection{The \texorpdfstring{\phifour}{phi4} model}

Given $n \ge 1$,
a \emph{spin field} is a function $\varphi : \Lambda_N \to \R^n$.
We write this function as $x \mapsto \varphi_x =(\varphi_x^1,\ldots,\varphi_x^n)$.

On $\R^n$, we use the Euclidean inner product $v \cdot w = \sum_{i=1}^n v^i w^i$,
the Euclidean norm $|v|^2 = v\cdot v$,
and write $|v|^4 = (v\cdot v)^2$.
Given $g>0$, $\nu \in \R$, we define a function $U_{g,\nu,N}$ of the field by
\begin{equation} \label{e:Vdef1}
  U_{g,\nu,N}(\varphi)
  = \sum_{x\in\Lambda}
  \Big(\tfrac{1}{4} g |\varphi_x|^4 + \half \nu |\varphi_x|^2
  + \half \varphi_x\cdot (-\Delta \varphi)_x  \Big)
  .
\end{equation}
Then the expectation of a random variable $F:(\R^n)^{\Lambda_N} \to \R$ is defined by
\begin{equation}
  \label{e:phi4-expectation-def}
  \langle F \rangle_{g,\nu,N}
  = \frac{1}{Z_{g,\nu,N}} \int F(\varphi) e^{-U_{g,\nu,N}(\varphi)} d\varphi,
\end{equation}
where $d\varphi$ is the Lebesgue measure on
$(\R^n)^{\Lambda}$,
and $Z_{g,\nu,N}$ is a normalisation constant (the \emph{partition function})
chosen so that $\langle 1 \rangle_{g,\nu,N} = 1$.
Given a lattice point $x$,
we define the finite and infinite volume \emph{two-point functions}
(whenever the infinite volume limit exists):
\begin{equation}\label{e:two-point-function-phi4}
G_{x, N}(g,\nu; n) =
\frac{1}{n} \pair{\varphi_0 \cdot \varphi_x}_{g,\nu, N},
\quad
G_x(g,\nu; n) = \lim_{N \to \infty} G_{x, N}(g,\nu; n).
\end{equation}
In the above limit, we identify a point $x \in \Zd$ with $x \in \Lambda_N$
for large $N$, by embedding the vertices of $\Lambda_N$ as an approximately
centred cube in $\Z^d$ (say as $[-\frac12 L^N+1,\frac12 L^N]^d \cap \Z^d$ if $L^N$ is even
and as $[-\frac12 (L^N-1), \frac12 (L^N-1)]^d \cap \Z^d$ if $L^N$ is odd).

\subsubsection{Weakly self-avoiding walk}

Let $X$ be the continuous-time simple random walk on the lattice $\Z^d$, with $d > 0$.
In other words, $X$ is the stochastic process with right-continuous sample paths that
takes steps uniformly at random to one of the $2d$ nearest neighbours of the current position
at the events of a rate-$2d$ Poisson process.
Steps are independent of the Poisson process and of all other steps.
Let $E_0$ denote the
expectation for the process
with $X(0)=0 \in \Zd$.
The \emph{local time} of $X$ at $x$ up to time $T$ is the random
variable $L_T(x) = \int_0^T \1_{X(t)=x} \; dt$,
and the \emph{self-intersection local time} up to time $T$ is the random variable
\begin{equation} \label{e:intersection-local-time-def}
  I(T) = \int_0^T \!\! \int_0^T \1_{X(t_1) = X(t_2)} \; dt_1 \, dt_2
  =
  \sum_{x\in\Z^d} \big(L_T(x)\big)^2.
\end{equation}

Given $g>0$, $\nu \in \R$, and $x \in \Zd$, the continuous-time weakly self-avoiding walk
\emph{two-point function} is defined by the (possibly infinite) integral
\begin{equation}
  \label{e:two-point-function-wsaw}
  G_x(g,\nu; 0)
  =
  \int_0^\infty
  E_0 \left(
    e^{-g I(T)}
    \1_{X(T)=x} \right)
  e^{- \nu T}
  dT.
\end{equation}
We write
$G_{x, N}$ for the finite volume analogue of \eqref{e:two-point-function-wsaw}
on the torus $\Lambda_N$.

\subsubsection{Critical point and correlation length of order \texorpdfstring{$p$}{p}}
\label{sec:clp}

For both models, i.e., for all integers $n\ge 0$, the \emph{susceptibility} is defined by
\begin{equation}
\label{e:susceptibility-def}
\chi(g, \nu; n)
= \lim_{N\to\infty} \sum_{x\in\Lambda_N} G_{x,N} (g,\nu; n).
\end{equation}
The limit exists for $n=0$ \cite{BBS-saw4-log},
but the general case is incomplete for $n \ge 1$ due to a lack of correlation inequalities
for $n >2$ \cite{FFS92}.
The existence of the limits
\refeq{two-point-function-phi4} and \refeq{susceptibility-def}
in the contexts we study is established in
\cite{BBS-phi4-log,ST-phi4}
(assuming $L$ is large).

We write $a \sim b$ to mean $\lim a/b=1$.
It is proved in \cite{BBS-phi4-log,BBS-saw4-log}
that for $n \ge 0$ and small $g>0$
there exists a \emph{critical value} $\nu_c=\nu_c(g;n) < 0$ such that
the susceptibility diverges according to the asymptotic formula
\begin{equation}
\label{e:susceptibility-asympt}
\chi(g,\nu_c+\varepsilon ;n) \sim
A_{g,n}\varepsilon^{-1}(\log \varepsilon^{-1})^\frac{n+2}{n+8}
\quad \text{as } \varepsilon\downarrow 0,
\end{equation}
for some amplitude $A_{g,n}>0$.
Also, in \cite{BBS-saw4-log,ST-phi4}, it is proved that
\begin{equation} \label{e:Gnuc-asympt}
  \lim_{\varepsilon\downarrow 0}\lim_{N\to\infty} G_{x,N}(g, \nu_c+\varepsilon; n)
  \sim
  (1+z_0^c(g;n)) (-\Delta_{\Z^4})^{-1}_{0x} \quad \text{as } |x|\to\infty,
\end{equation}
for some function $1+z_0^c(g;n)=1+O(g)$ (the \emph{field strength renormalisation}).

When $g = 0$, $\nu_c (0; n) = 0$ for all $n \geq 0$.
For the \emph{free} two-point function (as opposed to \emph{interacting} when $g > 0$),
we use $m^2\geq 0$ instead of $\nu \geq \nu_c=0$,
as we will extend \refeq{Gnuc-asympt} to approximate
$G_{x,N}(g, \nu; n)$
by its free counterpart $(1+z_0^c)G_{x,N}(0, m^2)$
with carefully chosen $m^2$.
The free two-point function is independent of $n\ge 0$, and its infinite-volume
version is equal to the lattice Green function
\begin{equation}
\label{e:Green-function}
G_x(0, m^2) = (-\Delta_{\Z^4} + m^2)^{-1}_{0x}.
\end{equation}
In probabilistic terms, $G_x(0, m^2)$ equals $\frac{1}{2d}$ times the expected number of
visits to $x$ of a simple random walk on $\Z^4$ with killing rate $m^2$, started from $0$.
It is proved in \cite{BBS-phi4-log, BBS-saw4-log} that for
$n \ge 0$ and for small $g > 0$,
$\nu_c(g;n) = -{\sf a} g + O(g^2)$ with
${\sf a}= (n+2) (-\Delta_{\Z^4})^{-1}_{00}> 0$.

Given a unit vector $e\in \Zd$,
the \emph{correlation length} $\xi$ is defined by
\begin{equation}\label{e:xidef}
\xi (g,\nu; n) = \limsup_{k \to \infty} \frac{k}{\log G_{ke}(g, \nu; n)}.
\end{equation}
It provides a characteristic length scale for the model.
We study a related quantity,
the \emph{correlation length of order} $\clo >0$,
defined in terms of the infinite volume two-point function and susceptibility by
\begin{equation}
\label{e:clp}
\xi_{\clo} (g, \nu; n)
=
    \left[\frac{\sum_{x \in \Z^4} |x|^\clo G_{x}(g, \nu; n)}
    {\chi(g, \nu; n)}\right]^\frac{1}{\clo}
\hspace{-8pt}.
\end{equation}
It is predicted that $\xi_p$ has the same asymptotic behaviour near $\nu = \nu_c$ as
the correlation length $\xi$, for all $p>0$.

For all quantities defined above, we often omit the argument $n$ from the notation.

\subsection{Main result}
\label{sec:mr}

Our main result is the following theorem.
We define constants ${\sf c}_p>0$ by
\begin{equation} \label{e:cpdef}
    {\sf c}_p^p = \int_\Rd |x|^p (-\Delta_\Rd + 1)^{-1}_{0x} \; dx,
\end{equation}
and set $\tilde A_{g,n} = A_{g,n}/(1+z_0^c)$ for the constants $A_{g,n}$ and $z_0^c$
in \eqref{e:susceptibility-asympt}--\eqref{e:Gnuc-asympt}.
(We expect that $A_{g,n},z_0^c$ are independent of $L$, but this has not been proved.)

\begin{theorem}\label{thm:mr}
Let $d=4$, $n \geq 0$ and $p>0$.
For $L$ sufficiently large (depending on $\clo,n$), and for
$g >0$ sufficiently small (depending on $\clo,n$),
as $\varepsilon \downarrow 0$,
\begin{equation}
\lbeq{xipasy}
\xi_\clo(g, \nu_c  + \varepsilon;n)
\sim {\sf c}_p \tilde A_{g,n}^{\frac12}  \varepsilon^{-\frac{1}{2}} (\log \varepsilon^{-1})^{\frac{1}{2}\frac{n+2}{n+8}}
.
\end{equation}
\end{theorem}

Some results related to Theorem~\ref{thm:mr} have been obtained previously.
For $n=1$, the $\varepsilon^{-\frac{1}{2}} (\log \varepsilon^{-1})^{\frac{1}{6}}$
behaviour on the right-hand side of \refeq{xipasy} was proven
in \cite{HT87} for the correlation
length $\xi$ of \refeq{xidef},
in the sense of upper and lower bounds with different constants.
For the $n=0$ model, the end-to-end distance of
a \emph{hierarchical} version of the continuous-time weakly self-avoiding walk, up to time $T$,
was shown to have $T^{\frac 12} (\log T)^{\frac 18}$ behaviour \cite{BI03c}.

The proof of Theorem~\ref{thm:mr} involves a modification of the renormalisation group
strategy used in \cite{BBS-saw4-log,BBS-saw4,BBS-phi4-log,ST-phi4} to analyse the
susceptibility and the critical two-point function.
That strategy is based on a multi-scale analysis using
a finite range decomposition of the covariance $(-\Delta + m^2)^{-1} = \sum_j C_j$.
The new ingredient in our proof
is to take better advantage of the decay of $C_j$ when $j$ exceeds
the \emph{mass scale} $j_m$ given by $L^{j_m} \approx m^{-1}$.
Using this decay, beyond the mass scale we obtain better control over the
two-point function than what was obtained in \cite{BBS-saw4,ST-phi4},
sufficient to analyse $\xi_p$ and to prove Theorem~\ref{thm:mr}.
It would be of interest to extend this, to seek the
further improvements that would be needed to
analyse the correlation length $\xi$.  Our new treatment leads to the simplification
that at scales beyond $j_m$ the large-field regulator $\tilde G_j$ used in
\cite{BBS-saw4-log,BBS-saw4,BBS-phi4-log,ST-phi4} becomes superfluous, and
the fluctuation-field regulator $G_j$ suffices.

\subsection{The non-interacting model}

An elementary ingredient in the proof of Theorem~\ref{thm:mr} is the following result for
the $g=0$ case, which is independent of $n\ge 0$.
For simplicity, we restrict attention to dimensions $d>2$, as only $d=4$ is used in this paper.

\begin{prop}\label{prop:Gab-free-moment-estimate}
For all dimensions $d>2$ and all $p>0$,
as $m^2 \downarrow 0$,
\begin{equation}
\label{e:Gab-free-moment-estimate}
\sum_{x\in\Zd} |x|^p G_{x}(0, m^2)
=
{\sf c}_p^p m^{-(p + 2)} (1 + O(m)),
\end{equation}
with ${\sf c}_p$ given by \eqref{e:cpdef}.
In particular, $\xi_p(0,\varepsilon) = {\sf c}_p \varepsilon^{-1/2}
(1+O(\varepsilon^{1/2}))$ as $\varepsilon \downarrow 0$.
\end{prop}

Proposition~\ref{prop:Gab-free-moment-estimate} is presumably well-known,
but since we have not found a proof in the literature,
we provide a proof in Appendix~\ref{app:free-moments}.
Note that this $g = 0$ case does not exhibit a logarithmic correction.


\section{Proof of main result}
\label{sec:proof-mr}

In this section, we state Proposition~\ref{prop:R},
an improvement on the results of \cite{ST-phi4} (this reference subsumes and extends the
results of \cite{BBS-saw4}), and show that
Theorem~\ref{thm:mr} is a consequence of Proposition~\ref{prop:R}.

The main conclusions of \cite{ST-phi4} are based
on a rigorous renormalisation group method.
The method uses an approximation of the interacting model
by the noninteracting one,
encoded by an $n$-dependent map
\begin{equation}
\lbeq{mgnuz}
    (g,\varepsilon) \mapsto (m^2,g_0,\nu_0,z_0)
\end{equation}
with domain $[0,\delta)^2$ (for some small $\delta > 0$).
Properties of this map are discussed briefly in \cite[Section~\ref{phi4-sec:changevariables}]{ST-phi4} where
further references to \cite{BBS-saw4-log} and \cite{BBS-phi4-log} are given.
In particular, the map \refeq{mgnuz} identifies the values of $m^2,z_0^c$ that
lead to the approximate equality $G_{x,N}(g, \nu; n)
\approx (1+z_0^c)G_{x,N}(0, m^2)$ discussed in Section~\ref{sec:clp}.

A key ingredient of the renormalisation group method is a flow of renormalised coupling constants.
The flow of the important coupling constant $g$ is well approximated by
the sequence $\gbar$ defined by
\begin{equation} \label{e:gbar}
  \gbar_{j+1}
  =
  \gbar_j
  -
  \beta_j  \gbar_j^{2}, \qquad \gbar_0 = g_0,
\end{equation}
where the coefficients $\beta_j = \beta_j(m^2) > 0$ are defined in \cite[\eqref{phi4-log-e:betadef}]{BBS-phi4-log}.
The $\beta_j$ obey $\beta_j(m^2) \approx \beta_j(0)$ for $j\leq j_m$ and $\beta_j(m^2) \approx 0$ for $j\geq j_m$,
where
\begin{equation}
\lbeq{jmdef}
j_m = \lfloor\log_{L} m^{-1}\rfloor
\end{equation}
is the \emph{mass scale}.
We are interested in small $m$ with $L$ fixed, so $j_m$ is large and positive.
It follows that $\gbar_j$ decays like $1/j$ for $j \leq j_m$ and is approximately constant for $j>j_m$.
See \cite[Section~\ref{phi4-log-sec:pt2}]{BBS-phi4-log} for details.

We estimate sums over $x\in \Z^4$
by dividing $\Z^4$ into shells $S_1 = \{x : |x| < \frac 12 L\}$ and, for $j \ge 2$,
$S_j = \{x : \frac 12 L^{j-1} \le |x| < \frac 12 L^{j}\}$.
The number of points in $S_j$ is bounded by $O(L^{4j})$.
We refer to the integer $j$ as a \emph{scale}.
Given $x\in \Z^4$, we define the \emph{coalescence scale}
to be the unique scale $j_x$ such that
\begin{equation}
   \label{e:Phi-def-jc}
    x \in S_{j_x +1}
   .
\end{equation}
Equivalently, $j_{x} = \max\{0,\lfloor \log_{L} (2 |x|)\rfloor\}$; this introduces a minor
notational clash with the mass scale $j_m$ defined in \refeq{jmdef}
that should not cause problems.
It follows from \cite[Proposition~\ref{log-prop:approximate-flow}]{BBS-saw4-log} that
\begin{equation} \label{e:gjxgjmbd}
  \gbar_{j} = O((\log m^{-1})^{-1}) \;\; \text{for $j \geq j_m$}, \quad
  \gbar_{j_x} = O((\log |x|)^{-1}) \;\; \text{for $j_x \leq j_m$.}
\end{equation}

In \cite[Remark~\ref{phi4-rmk:massive-Gab}]{ST-phi4},
a remainder $R_x$ is identified such that
\begin{equation}\label{e:Gab-to-sum-Rqj}
\frac{1}{1 + z_0}G_{x}(g, \nu; n)
= \big(1 + O(\gbar_{j_{x}}) \big) G_{x}(0, m^2) + R_x(m^2,g_0;n),
\end{equation}
for any $n \geq 0$ with $m^2,g_0$ given in terms of $(g,\nu)$ by \refeq{mgnuz},
with
\begin{equation}
\lbeq{badRbd}
|R_{x}|
\le O(\gbar_{j_{x}}) G_{x}(0,0).
\end{equation}
Thus \refeq{Gab-to-sum-Rqj} compares the value of the interacting theory on the left-hand side,
evaluated at $(g,\nu)$, with the first term
on the right-hand side.  The first term on the right-hand side is the corresponding
free quantity at \emph{renormalised} parameter values $(0,m^2)$.

However, with \refeq{badRbd}, the exponential decay present in $G_{x}(0, m^2)$
when $m^2>0$
is overwhelmed by the remainder term which involves instead the massless free two-point
function $G_{x}(0, 0)$, and control needed for the correlation length of order $p$
gets lost.
In the next proposition,
we improve the estimate \refeq{badRbd} of \cite[Lemma~\ref{phi4-lem:qflow}]{ST-phi4}
by providing a new factor $(m|x|)^{-2s}$
when $|x|$ is large compared to the mass.
Roughly, $L^{j_x} \approx |x|$ and $L^{j_m} \approx m^{-1}$, so when the coalescence scale
exceeds the mass scale, $m|x|$ becomes greater than $1$.
Thus the factor $(m|x|)^{-2s}$
gives good decay when the coalescence scale exceeds the mass scale,
and we are free to choose $s > 0$ to be as large as desired.

\begin{prop}\label{prop:R}
Let $d=4$, $n \ge 0$, $\varepsilon \in (0,\delta)$ with $\delta$ sufficiently small,
and $\nu = \nu_c + \varepsilon$.
Let $x \in \Z^4$ with $x \neq 0$.
Fix any $s \geq 0$.
For $L$ sufficiently large and for $g > 0$ sufficiently small (depending on $s$),
\begin{equation}\label{e:Rab-bound}
|R_{x}|
\le
\frac{ O(\gbar_{j_{x}}) }{|x|^2}
\times
\begin{cases}
    1 & (m|x|\le 1)
    \\
    (m|x|)^{-2s} & (m|x|\ge 1),
\end{cases}
\end{equation}
with the  constant depending on $L$ and $s$.
\end{prop}

The proof of Proposition~\ref{prop:R} constitutes the main part of this paper
and is given in Sections~\ref{sec:Rpf1}--\ref{sec:Rpf2}.

The case $s=0$ of Proposition~\ref{prop:R}
is already given by \refeq{badRbd}.
This case is insufficient to prove Theorem~\ref{thm:mr}, as the remainder term $R_x$ is
not summable over $x \in \Z^4$ when $s=0$.
The improvement to arbitrary $s>0$ in  \eqref{e:Rab-bound} represents the key
innovation in this paper.
Note that, in particular, $R_x$
is summable after multiplication by $|x|^p$, provided $2s > p + 2$.

Before proving Proposition~\ref{prop:R}, we
prove Theorem~\ref{thm:mr} assuming Proposition~\ref{prop:R}.
In the proof, we use the important relation
that
\begin{equation}
\label{e:mass-epsilon-asympt}
 m^2
 \sim \tilde A_{g,n}^{-1} \varepsilon (\log \varepsilon^{-1})^{-\frac{n + 2}{n + 8}}
\quad \text{as $\varepsilon \downarrow 0$},
\end{equation}
which is proved in \cite[\eqref{phi4-log-e:masymp}]{BBS-phi4-log} for $n\geq 1$
and \cite[\eqref{log-e:m-mu}]{BBS-saw4-log} for $n=0$.
In particular, the dependence of $m^2$ on $\varepsilon$
encompasses the logarithmic correction for the susceptibility since
\begin{equation}\label{e:susceptibility-mass-identity}
\chi(g,\nu; n) = \frac{1 + z_0}{m^2},
\end{equation}
according to \cite[\eqref{phi4-log-e:chim2}]{BBS-phi4-log} for $n\geq 1$
and \cite[\eqref{log-e:chi-mtil0}]{BBS-saw4-log} for $n=0$.

\begin{proof}[Proof of Theorem~\ref{thm:mr}]
We multiply \eqref{e:Gab-to-sum-Rqj} by $|x|^p$, sum over $x \in \Z^4$,
and use \eqref{e:susceptibility-mass-identity},
to obtain
\begin{align}
\xi_p^p(g,\nu)
&=
\sum_{x \in \Z^4} |x|^p \frac{G_{x}(g, \nu)}{\chi(g, \nu)}
= m^2 \sum_{x \in \Z^4} |x|^p \Big(G_{x}(0, m^2) + r_{x}(g,m^2) \Big),
\end{align}
with
\begin{equation}
\lbeq{rx}
    r_x = O(\gbar_{j_x})  G_x(0, m^2) + R_{x}.
\end{equation}
By
Proposition~\ref{prop:Gab-free-moment-estimate},
this gives (as $m^2 \downarrow 0$)
\begin{align}
\lbeq{ximasy}
\xi_p^p(g,\nu)
    & \sim
    {\sf c}_p^p m^{-p} +
    m^2 \sum_{x \in \Z^4} |x|^p  r_{x}(g,m^2).
\end{align}
By \eqref{e:mass-epsilon-asympt}, it suffices to prove that
the first term on the right-hand side of \refeq{ximasy} is dominant.

For the term $O(\gbar_{j_x}) G_x(0, m^2)$ in \refeq{rx},
we apply \eqref{e:gjxgjmbd} to obtain
\begin{align}
\lbeq{easyerror}
    &\sum_{x \in \Z^4} \gbar_{j_x} |x|^p G_x(0, m^2)
    \nnb & \quad \le
    \sum_{x :  0<  j_x \le j_m} \frac{c |x|^p}{\log |x|} G_x(0, m^2)
    +
    \frac{c}{\log m^{-1}} \sum_{x : j_x >j_m}  |x|^p G_x(0, m^2).
\end{align}
In the first term,
we use $G_x(0, m^2) \le G_x(0, 0) \le O(|x|^{-2})$.  The restriction
$j_x \le j_m$ ensures that $|x| \le O(m^{-1})$.
Therefore the first term is bounded above by a multiple of
$(m^{-1})^{d+p-2}(\log m^{-1})^{-1}$, which suffices.
For the term with $j_x > j_m$, we extend the sum to $x \in \Z^4$
and apply Proposition~\ref{prop:Gab-free-moment-estimate}
to obtain a bound of the same form as for the first term.

Fix any $s>\frac 12 (p+2)$.
For the term $R_x$ of \eqref{e:rx}, we use Proposition~\ref{prop:R}
to see that
\begin{equation}\label{e:rab-bound-scales-bis}
|R_x(g, m^2)|
=
O(\gbar_{j_x})
L^{-2j_x - 2s (j_x - j_m)_+}.
\end{equation}
By \eqref{e:rab-bound-scales-bis} and \refeq{Phi-def-jc},
\begin{equation} \label{e:xpr-shells}
\begin{aligned}
    \sum_{x\in\Z^4} |x|^p |R_{x}(g,m^2)|
    &= \sum_{j = 1}^\infty\sum_{x\in S_j}   |x|^p |R_{x}(g,m^2)|
    \\
    &= \sum_{j = 1}^\infty
    L^{4j + pj - 2j - 2s (j - j_m)_+} O(\gbar_{j})
    ,
\end{aligned}
\end{equation}
with an $L$-dependent constant.
By Lemma~\ref{lem:mass-scale-sum} below (with $a=p+2$ and $b=1$),
we obtain
\begin{equation}\label{e:xpr-mbound}
    m^2 \sum_{x\in\Z^4} |x|^p |R_{x}(g,m^2)|
    = O\big( m^{-p}(\log m^{-1})^{-1}\big).
\end{equation}
The first term on the right-hand side of \refeq{ximasy} therefore dominates,
and the proof is complete.
\end{proof}

The estimate used to obtain \eqref{e:xpr-mbound}
is given by the following lemma,
which is stated more generally for use
in the proof of Proposition~\ref{prop:Gab-free-moment-estimate}.

\begin{lemma} \label{lem:mass-scale-sum}
Let $L>1$, $2s> a > 0$, $b \geq 0$, and let $\gbar_0>0$ be sufficiently small.
Then
\begin{equation} \label{e:gmsumbd}
\sum_{j=1}^\infty L^{aj - 2s (j - j_m)_+}
\gbar_j^b = O(m^{-a} \gbar_{j_m}^b) = O(m^{-a} (\log m^{-1})^{-b}).
\end{equation}
\end{lemma}

\begin{proof}
We divide the sum at the mass scale as
\begin{equation} \label{e:xpr-2sums}
\sum_{j=1}^\infty L^{aj - 2s (j - j_m)_+} \gbar_j^{b}
= \sum_{j=1}^{j_m} L^{aj} \gbar_j^{b} +  \sum_{j=j_m+1}^\infty L^{aj - 2s (j - j_m)} \gbar_{j}^{b}.
\end{equation}
For the second sum on the right-hand side, we use $\gbar_j = O(\gbar_{j_m})$ for $j > j_m$
by \eqref{e:gjxgjmbd},
and obtain
a bound consistent with the first equality of \refeq{gmsumbd}.
For the first term, we use the crude bound
$\gbar_i/\gbar_{i+1} = 1+O(g_0)$ (by
\cite[Lemma~\ref{flow-lem:elementary-recursion}]{BBS-rg-flow}), and find
\begin{equation}
  \sum_{j=1}^{j_m} L^{aj} \gbar_j^b
  \leq
  L^{aj_m} \gbar_{j_m}^b
  \sum_{j=1}^{j_m} ((1+O(\gbar_0))L^{-a})^{j_m-j}
  =
  O(L^{aj_m} \gbar_{j_m}^b),
\end{equation}
for sufficiently small $\gbar_0>0$.
This proves the first equality in \eqref{e:gmsumbd}.
The second equality then follows since
$\gbar_{j_m} = O(\log m^{-1})$ by \eqref{e:gjxgjmbd}.
\end{proof}


\section{Renormalisation group method}
\label{sec:rg}

We now provide a brief
outline of the renormalisation group method developed in
\cite{BS-rg-norm,BS-rg-loc,BBS-rg-pt,BS-rg-IE,BS-rg-step},
and identify the changes in the analysis of \cite{ST-phi4} needed
to prove Proposition~\ref{prop:R}.
We discuss some of the key points which have been explained in
detail elsewhere, and make no attempt at completeness here.
For notational simplicity,
we consider the case $n \geq 1$ below; the case $n = 0$ is similar.

\subsection{The two-point function via Gaussian integration}

The starting point of our study is the two-point function $G_{x,N}(g, \nu; n)$.
The first step is a change of variables.
Given $g>0,\nu\in \R$, and  given $m^2>0$ and $z_0 >-1$, let
\begin{equation}
  \label{e:gg0}
  g_0 = g(1+z_0)^2, \quad \quad
  \nu_0 = (1+z_0)\nu-m^2
  .
\end{equation}
Let $C = (-\Delta + m^2)^{-1}$ and let $\Ex_C$ denote the expectation with respect to
the Gaussian measure with covariance $C$.
For $y \in \Lambda$, we define the monomials
\begin{equation}
\label{e:tauphi}
\tau_y = \half |\varphi_y|^2,
\quad \tau_{\Delta,y} = \half \varphi_y \cdot (-\Delta \varphi)_y.
\end{equation}
Let ${\sf h} = n^{-1/2}(1, \ldots, 1) \in \R^n$.
With the two points $0,x\in \Lambda$ fixed,
we introduce \emph{observable fields} $\sigmaa, \sigmab \in \R$, and define
\begin{align}
    \label{e:V0def}
    U_0(\Lambda)
    &=
    \sum_{y\in\Lambda}
    \left(  g_{0} \tau_y^2 +   \nu_{0} \tau_y +  z_{0} \tau_{\Delta,y} \right)
    -
    \sigma_\pp (\varphi_{\pp} \cdot {\sf h})
    - \sigma_\qq (\varphi_{\qq} \cdot {\sf h}),
\end{align}
and
\begin{align}
\lbeq{Z0def}
    Z_0 = e^{-U_0(\Lambda)}.
\end{align}
It is then an elementary calculation (see \cite[\eqref{phi4-e:DaDbPN}]{ST-phi4}) to show
that (for $n \geq 1$)
\begin{align}
\label{e:generating-fn}
    G_{x,N}(g, \nu; n)
    &=
    (1+z_0)
    \frac{\partial^2 }{\partial \sigma_\pp  \partial \sigma_\qq }
    \Big|_{0}
    \log
    \Ex_C  Z_0.
\end{align}

The renormalisation group method provides a way to calculate the integral
$\Ex_C Z_0$ and thereby compute \eqref{e:generating-fn}.
At this point, $z_0$ and $m^2$ are arbitrary, but careful choices of parameters will
be required to make \refeq{generating-fn} useful, as in \refeq{mgnuz}, and it is part
of the method to determine this careful choice.

\subsection{Progressive integration}

We evaluate the Gaussian integral $\Ex_C Z_0$ progressively, via
the covariance decomposition
\begin{equation}
\label{e:NCj}
C = C_1 + \cdots + C_{N-1} + C_{N,N}
\end{equation}
constructed in \cite{Baue13a} (see also \cite{BGM04}). For simplicity, we write $C_N = C_{N,N}$.
For an integrable function $F$ of the spin field $\varphi$, we let
$\Ex_{w}\theta F$ be the convolution of $F$ with the Gaussian measure of covariance $w$, i.e.,
$(\Ex_w\theta F)(\varphi) = \Ex_wF(\varphi + \zeta)$ where the expectation integrates the
variable $\zeta$.
It is a property of Gaussian integration (see \cite{BS-rg-norm}) that
\begin{equation}
    (\Ex_C \theta F)(\varphi)
    =
    (\Ex_{C_N}\theta \circ \Ex_{C_{N-1}}\theta \circ \ldots \circ \Ex_{C_1}\theta F)
    (\varphi).
\end{equation}
Let \begin{equation}
Z_N = \Ex_C \theta Z_0
=
\Ex_{C_N}\theta \circ \Ex_{C_{N-1}}\theta \circ \ldots \circ \Ex_{C_1}\theta Z_0.
\end{equation}
In particular,
\begin{equation}
\Ex_C Z_0 = Z_N(0).
\end{equation}
This allows us to evaluate the integral $\Ex_C Z_0$ by studying the
dynamical system $Z_j \mapsto Z_{j+1}$ defined by
\begin{equation}
Z_{j+1} = \Ex_{C_{j+1}} \theta Z_j, \quad j < N.
\end{equation}

For its analysis, we require
a suitable space $\Ncal$ of functions of the spin
and observable fields, on which the dynamical system acts.  The space $\Ncal$ is
discussed in detail in \cite[Section~\ref{phi4-sec:phi4observables_representation}]{ST-phi4}.
The part of $\Ncal$ which does not involve the observable fields $\sigmaa,\sigmab$ is
given by
\begin{equation}
\label{e:Ncaldef}
    \Ncal^\varnothing = \Ncal^\varnothing(\Lambda) = C^{p_\Ncal}((\R^n)^\Lambda,\R).
\end{equation}
The finite smoothness parameter $p_\Ncal$ is discussed in Section~\ref{sec:newnorm} below,
where it is explained that $p_\Ncal$ must be chosen in a way that depends on the
parameter $p$ in Theorem~\ref{thm:mr}.
The part of $\Ncal$ involving the observable fields contains some subtleties that
need not concern us here; see \cite[Section~\ref{phi4-sec:phi4observables_representation}]{ST-phi4}
for details.

\subsection{Local field polynomials}

The dynamical system is analysed via a perturbative part which is tracked accurately
to second order in $g$, together with a third-order non-perturbative part whose study
forms the main part of our effort.  For the perturbative part, we first introduce
an appropriate space of local field polynomials.

For $y \in \Lambda$, we supplement \refeq{tauphi} by defining
\begin{equation}
\label{e:tauphi2}
\quad \tau_{\nabla\nabla,y}
= \frac{1}{4} \sum_{e\in\Z^d:|e| = 1} \nabla^e \varphi_y \cdot \nabla^e \varphi_y.
\end{equation}
With $x \in \Lambda$ fixed, and
given $g,\nu,z,y,u,\lambda_0,\lambda_x,q_0,q_x \in \R$, we extend
\refeq{V0def}
by defining the polynomial
\begin{align}
    V_y &= g \tau_y^2 + \nu \tau_y + z \tau_{\Delta,y} + y \tau_{\nabla\nabla,y} + u
    \nnb
    & \quad
    - \1_{y=\pp}\lambda_{\pp}(\varphi_{\pp} \cdot {\sf h})\sigma_\pp
    - \1_{y=\qq}\lambda_{\qq}(\varphi_{\qq} \cdot {\sf h})\sigma_\qq
     \nnb
    & \quad
    - \textstyle{\frac 12} (\1_{y=\pp} q_\pp + \1_{y=\qq}q_\qq )\sigma_\pp\sigma_\qq .
\lbeq{Vy}
\end{align}
Then we define $\Vcal$ to be the space of functions $V=V_y$ of the form \refeq{Vy}.
Given $X \subset \Lambda$, we also define
\begin{equation}
\label{e:Vcalesig}
    \Vcal(X) = \{V(X) = \textstyle{\sum_{y\in X}} V_y : V \in \Vcal \}.
\end{equation}

We also make use of the subspaces $\Vcal^{(1)} \subseteq \Vcal$ consisting of polynomials with $y = 0$, as well as the subspace
$\Vcal^{(0)} \subseteq \Vcal^{(1)}$ of polynomials with
$u = y =   q_\pp=q_\qq = 0$.
For $V \in \Vcal$, we define maps $V \mapsto V^{(1)} \in \Vcal^{(1)}$
and $V \mapsto V^{(0)} \in \Vcal^{(0)}$. Both maps replace
$z\tau_{\Delta}+y\tau_{\nabla\nabla}$ by
$(z+y)\tau_{\Delta}$, and the latter
additionally sets
$u = q_\pp = q_\qq = 0$.

\subsection{Renormalisation group coordinates}
\label{sec:rgcoord}

For $j=0,\ldots, N$,
we partition $\Lambda$ into $L^{N-j}$ disjoint scale-$j$ blocks of side length $L^j$.
A scale-$j$ \emph{polymer} is a union of scale-$j$ blocks.
The set of all scale-$j$ blocks is denoted $\Bcal_j$, and
the set of all scale-$j$ polymers is denoted $\Pcal_j$.
For $X \in \Pcal_j$, we write $\Bcal_j(X)$ for the set of scale-$j$ blocks in $X$.
For $F, G : \Pcal_j \to \Ncal$, we define the \emph{circle product} $F \circ G : \Pcal_j \to \Ncal$ by
\begin{equation}
(F \circ G)(X) = \sum_{Y\in\Pcal_j(X)} F(X \setminus Y) G(Y).
\end{equation}

The evolution of $Z_j$ can be tracked in the \emph{renormalisation group coordinates}
$\zeta_j \in \R$,
$I_j, K_j : \Pcal_j \to \Ncal$, defined such that
\begin{equation}
\label{e:IcircKnew}
    Z_j = e^{\zeta_j}(I_j\circ K_j)(\Lambda),
    \qquad
    \zeta_j= - u_j|\Lambda|
    + \textstyle{\frac 12} (q_{\pp,j} + q_{\qq,j}) \sigma_\pp\sigma_\qq
    .
\end{equation}
The coordinate $I_j$ tracks the evolution of the
\emph{relevant} and \emph{marginal} directions.  It
is determined by a local polynomial
$U\in \Vcal^{(0)}$,
and takes the form
\begin{equation}
I_j(X) = \prod_{B \in \Bcal_j(X)} e^{-U(B)} (1 + W_j(B, U)), \quad X \in \Pcal_j,
\end{equation}
with $W_j$ an explicit quadratic term in $U$ (defined in \cite[\eqref{pt-e:WLTF}]{BBS-rg-pt}).
The evolution of $(\zeta, U)$ to second order is called the \emph{perturbative flow} and is
given by the explicit map $\Vpt : \Vcal \to \Vcal$ defined in
\cite[\eqref{pt-e:Vptdef}]{BBS-rg-pt}.
In particular, it is shown in \cite[Proposition~\ref{phi4-prop:pt}]{ST-phi4}
that the perturbative flow of $q$ is given by
\begin{align}
\label{e:qpt}
q_\pt = q + \lambda_0 \lambda_x C_{j+1;0x},
\end{align}
and that the perturbative flow of $\lambda_0$ and $\lambda_x$ becomes the identity map
once $j$ exceeds the coalescence scale $j_x$.

At scale $j = 0$, we are given $U_0$
as defined in \eqref{e:V0def} and we set $\zeta_0 = 0$.
In particular,
the initial values of $u,q_0,q_x$ are zero, and the initial values of $\lambda_0,\lambda_x$
are $1$.
By definition, $W_0 = 0$, and we have $I_0(X) = e^{-U_0(X)}$.
We define $\1_\varnothing : \Pcal_0 \to \Ncal$ by
\begin{equation}
\1_\varnothing(X) =
\begin{cases}
1 & X = \varnothing \\
0 & \text{otherwise},
\end{cases}
\end{equation}
and set $K_0 = \1_\varnothing$.
With these choices, $Z_0$ of \refeq{Z0def}
takes the form \eqref{e:IcircKnew}, and we seek $(\zeta_j, U_j, K_j)$ such that
this continues to hold as the scale advances.

Equivalently, given $(U_j, K_j)$, we define $(\delta\zeta_{j+1}, U_{j+1}, K_{j+1})$ so that
\begin{equation} \label{e:IcircKdu}
  \Ex_{j+1}\theta(I_j \circ K_j)(\Lambda)
  =
  e^{-\delta \zeta_{j+1}}(I_{j+1} \circ K_{j+1})(\Lambda),
\end{equation}
where $\delta\zeta_{j+1} = \zeta_{j+1} - \zeta_j$.
Moreover, we need $K_j$ to contract as the scale advances, under an appropriate norm.
The construction of (scale-dependent) maps $V_+$ and $K_+$ such that
\eqref{e:IcircKdu} holds with
\begin{equation}
    (\delta\zeta_{j+1}, U_{j+1}, K_{j+1}) = (V_+(U_j, K_j), K_+(U_j, K_j))
\end{equation}
is the main accomplishment of \cite{BS-rg-step} and is summarised
in Section~\ref{sec:step} below, in a form adapted to our current setting.

\subsection{The main theorem}
\label{sec:step}

We define a scale-dependent norm
\begin{equation}
\label{e:Vnormdef}
\begin{aligned}
\|V\|_{\Vcal} &=
\max\Big\{
|g|, L^{2j}|\nu|, |z|, |y|,  L^{4j}|u|,
\ell_j\ell_{\sigma,j}(|\lambda_\pp|\vee|\lambda_\qq|),\;
 \ell_{\sigma,j}^{2} (|q_\pp|\vee|q_\qq|)
\Big\}
\end{aligned}
\end{equation}
on $V \in \Vcal$, which depends on parameters $\ell_j$ and $\ell_{\sigma,j}$.
An innovation in this paper is that we define these parameters by
\begin{align}
\label{e:elldef-zz}
\ell_j &= \ell_0 L^{-j - s (j - j_m)_+}, \quad
\ell_{\sigma,j}
=
\ell_{j \wedge j_{x}}^{-1} 2^{(j - j_{x})_+} \ggen_j,
\end{align}
where the mass scale $j_m$ is defined in \eqref{e:jmdef},
the coalescence scale
$j_x$ is defined in \eqref{e:Phi-def-jc},
and $s$ is the parameter appearing in Proposition~\ref{prop:R}.
The sequence $\ggen = \ggen(m^2,g_0)$ is defined in
\cite[\eqref{log-e:ggendef}]{BBS-saw4-log};
it is bounded above and below by constant multiples of
the sequence $\gbar$ defined in
\eqref{e:gbar},
by
\cite[Lemma~\ref{log-lem:gbarmcomp}]{BBS-saw4-log}.
We discuss the origin of the definition \refeq{elldef-zz} in detail
in Section~\ref{sec:Rpf1}.

The following theorem is a restatement of \cite[Theorem~\ref{phi4-thm:step-mr-fv}]{ST-phi4}
with three changes. The first, minor, change is the
specialisation to the case $p = 1$ and $h = {\sf h}$
(in the notation of \cite{ST-phi4}). The second change is the
main accomplishment of this paper, namely that the norms in the estimates
\eqref{e:RKplus} below use the new norm parameters \eqref{e:elldef-zz}.
The third change is that we have omitted some technical details
concerning the parameter $m^2$ to simplify this brief summary;
these details are as in \cite[Theorem~\ref{phi4-thm:step-mr-fv}]{ST-phi4}.
In particular, $m^2$ must be chosen small in Theorem~\ref{thm:step-mr-fv}.

In \cite{BS-rg-step}, maps $V_+,K_+$ are defined which map a pair $(U,K)$ at scale $j$
to $(V_+(U,K),K_+(U,K))$ at scale $j+1$, and which preserve the circle product
$I\circ K$ under expectation as in \refeq{IcircKdu}.
A norm has already been defined on the space $\Vcal$ in \refeq{Vnormdef}.
We also require a norm on a space $\Kcal$ containing the non-perturbative
coordinate $K$ (see \cite[Definition~\ref{phi4-def:Kspace}]{ST-phi4}),
which is the $\Wcal_j$ norm of \cite[\eqref{step-e:9Kcalnorm}]{BS-rg-step}.
We denote the ball of radius $r$ in
the normed space $\Wcal_j$ by $B_{\Wcal_j}(r)$.
Given $C_\DV>0$ and $\alpha>0$, we define the domains
\begin{align}
\label{e:DVdef}
    \DV_j &= \{U\in \Vcal^{(0)} :
    g> C_{\DV}^{-1} \ggen_j  , \;  \|U\|_{\Vcal} < C_{\DV} \ggen_j \},
\\
\domRG_j &= \DV_j \times B_{\Wcal_j}(\alpha \chigen_j \ggen_j^3),
\end{align}
where $\ggen_j$ was discussed above,
and $\chigen_j$ is a sequence that, roughly, is equal to 1 below the mass scale and decays
exponentially above the mass scale (discussed above \cite[\eqref{step-e:domRGgen}]{BS-rg-step}).
Then, at scale $j$, the maps $V_+,K_+$ act on the domain $\domRG_j$
and map into $\Vcal_{j+1}^{(1)},\Kcal_{j+1}$, respectively.
The deviation of the map $V_+$ from the perturbative map $\Vpt$ (mentioned above \refeq{qpt})
is denoted by $R_+$, and is defined by
\begin{equation}
\label{e:Rplusdef}
    R_+(V,K) = V_+(V,K) -\Vpt^{(1)}(V).
\end{equation}
The following theorem is applied with $\alpha =4M$ as a convenient choice.

\begin{theorem}
\label{thm:step-mr-fv}
Let $d = 4$ and let $n \ge 0$. Fix $s > 0$.
Let $C_\DV$ and $L$ be sufficiently large.
There exist $M>0$ and $\delta >0$ such that
for $\ggen \in (0,\delta)$, 
and with the domain
$\domRG$ defined using any $\DVa> M$, the maps
\begin{equation}
\label{e:RKplusmaps}
R_+:\domRG 
\to \Vcal^{(1)},
\quad
K_+:\domRG 
\to \Wcal_{+}
\end{equation}
define $(U,K)\mapsto (V_+,K_+)$ obeying \eqref{e:IcircKdu},
and satisfy the estimates
\begin{equation}
\label{e:RKplus}
\|R_+\|_{\Vcal}
\le
M\chigen_+\ggen_+^{3}
, \qquad
\|K_+\|_{\Wcal_+}
\le
M\chigen_+ \ggen_+^{3}
.
\end{equation}
\end{theorem}

The proof of Theorem~\ref{thm:step-mr-fv} is identical to the proof of
\cite[Theorem~\ref{phi4-thm:step-mr-fv}]{ST-phi4}, via a version of
\cite[Theorems~\ref{step-thm:mr-R}--\ref{step-thm:mr}]{BS-rg-step} that
uses the norm parameters \eqref{e:elldef-zz} with $s > 0$.
The proof of the latter results with these new norm parameters amounts to
checking that the proof of the $s = 0$ case contained in \cite{BS-rg-IE,BS-rg-step}
continues to hold with $s > 0$. A verification of this fact is carried
out in Section~\ref{sec:Rpf2} below.

Theorem~\ref{thm:step-mr-fv} expresses a contractive property of the map $K_+$,
as it takes $K$ in a ball whose radius involves $\alpha=4M$ at scale $j$ to
an image which lies in a ball whose radius involves the smaller number $M$ at scale
$j+1$.  The fact that $K_+$ is a contraction is used
in \cite[Proposition~\ref{log-prop:KjNbd}]{BBS-saw4-log} (for $n=0$) and
\cite[Theorem~\ref{phi4-log-thm:flow-flow}]{BBS-phi4-log}
(for $n \ge 1$) to prove that, for $m^2$ and
$g_0$ sufficiently small, there exist \emph{critical} initial conditions
$\nu_0 = \nu_0^c(m^2, g_0)$ and $z_0 = z_0^c(m^2, g_0)$ such that,
for the case of no observables ($\sigma_0=\sigma_x=0$), iteration of
the maps $(V_+,K_+)$ defines a
sequence $(V_j^{(0)}, K_j)$ which lies
in the domain $\domRG_j$ and obeys the estimates \refeq{RKplus}
\emph{for all} $j = 1, \ldots, N$.
This construction of critical initial conditions uses the $s=0$ version
of \refeq{elldef-zz}.

The case with observable fields included is handled in \cite{ST-phi4}.
Because we have increased $\ell_{\sigma,j}$ beyond the mass scale, the
estimates on $q_0,q_x$ given by the bound on $R_+$ in \refeq{RKplus} are
significantly improved compared to their versions with $s=0$ in \cite{ST-phi4}.
As is discussed in detail in  \cite[Section~\ref{phi4-sec:pfmr1}]{ST-phi4},
$V_j^{(0)}$ remains in the domain $\domRG_j$ for all $j$
(also concerning $\lambda_{0,j}, \lambda_{x,j}$).
Moreover,
$\lambda_{0,j}, \lambda_{x,j},q_{0,j},q_{x,j}$, are independent of the volume parameter $N$ and
so can be extended to infinite sequences, and the following limits exist:
\begin{equation}
q_{u,\infty} = \lim_{j\to\infty} q_{u,j}, \quad u = 0, x.
\end{equation}

\subsection{Identity for the two-point function}
\label{sec:approxG}

At scale $N$ the torus $\Lambda$ is a single block, and  \eqref{e:IcircKnew} gives
\begin{equation}
Z_N = e^{\zeta_N} (I_N(\Lambda) + K_N(\Lambda)).
\end{equation}
Evaluation at $\varphi = 0$ gives
\begin{equation}
\label{e:ZN0}
Z_N(0) = e^{\zeta_N} (1 + K_N(\Lambda, 0)).
\end{equation}
Thus, by \eqref{e:generating-fn}
\begin{equation}
\lbeq{GK}
\frac{1}{1+z_0} G_{x,N}(g, \nu)
= \frac{1}{2} (q_{0,N} + q_{x,N})
+ \frac{D^2_{\sigma_a\sigma_b}K_{N}}{1 + K_{N}}
- \frac{\left(D_{\sigma_a}K_{N}\right)
\left(D_{\sigma_b}K_{N}\right)}{(1 + K_{N})^2}
\end{equation}
(with derivatives taken at $\sigma_0=\sigma_x=0$ on the right-hand side).
The bound \cite[\eqref{phi4-e:Kg1}]{ST-phi4} is only improved by the new choice of norm,
to assert that, for $l=0,1,2$, the $l^{\rm th}$ derivative of $K_N(\Lambda)$
with respect to the observable field is
bounded above by a multiple of $2^{-l(N-j_x)}|x|^{-l} L^{-ls(N-j_x)} \chicCov_N \gbar_N^{3-l}$.
In particular, the last two terms of \refeq{GK} vanish as $N \to \infty$, and
\begin{equation}
\frac{1}{1+z_0} G_x(g, \nu; n) = \frac{1}{2} (q_{0,\infty} + q_{x,\infty}).
\end{equation}

It is then a consequence of
\eqref{e:qpt} and \eqref{e:Rplusdef} (as in the proof of \cite[Lemma~\ref{phi4-lem:qflow}]{ST-phi4}) that
\begin{equation}
\label{e:q}
q_{u,\infty}(m^2)
= \lambda_{\pp, j_\qq} \lambda_{\qq, j_\qq}  G_x(0, m^2) + \sum_{i = j_\qq}^\infty R^{q_u}_i,
\quad u = 0, x,
\end{equation}
where $R^{q_u}_i$ is the coefficient of $\1_{y=u}\sigma_0\sigma_x$ (recall \refeq{Vy})
of $R_{+,i}$ (recall \refeq{Rplusdef}).
Moreover, as in \cite[\eqref{phi4-e:lam-star}]{ST-phi4} and \cite[Corollary~\ref{phi4-cor:vx}]{ST-phi4},
\begin{equation}
\lambda_{u,j_\qq} = 1 + O(\chicCov_{j_\qq} \gbar_{j_\qq}).
\end{equation}
It follows that
\begin{equation}
\frac{1}{1+z_0} G_x(g, \nu; n) = (1 + O(\gbar_{j_x})) G_x(0, m^2) + R_x,
\end{equation}
with
\begin{equation}
\label{e:Rabdef}
R_x = \frac{1}{2} \sum_{i=j_\qq}^\infty (R^{q_0}_i + R^{q_x}_i).
\end{equation}
This provides a more detailed statement of \eqref{e:Gab-to-sum-Rqj}.

\subsection{Proof of Proposition~\ref{prop:R}}
\label{sec:R-proof}

By the first bound of \eqref{e:RKplus} and the definition \eqref{e:Vnormdef},
\begin{align}
\label{e:vq-new}
    |R^{q_u}_{+,i}|
&
\le O(\ell_{\sigma,i}^{-2}\chicCov_i\gbar_i^{3}).
\end{align}
Using the old norm parameters $\ell_\sigma^\old$, the sum $R_x$ in \eqref{e:Rabdef}
is bounded by the right-hand side of \eqref{e:badRbd}.
With the new norm parameters, we instead get the result of Proposition~\ref{prop:R}.

\begin{proof}[Proof of Proposition~\ref{prop:R}
(assuming Theorem~\ref{thm:step-mr-fv})]
We insert the definition of $\ell_{\sigma,j}$ from \refeq{elldef-zz} into \refeq{vq-new}.
We also use $\ggen_j^{-2} = O(\gbar_j^{-2})$, $\chicCov_i \le 1$, $\ell_0^2 \le O(1)$,
as well as $\gbar_{j} \leq O(\gbar_{j_{x}})$ for $j \geq j_x$.
The definitions of
the coalescence scale $j_x$ and the mass scale $j_m$ imply that  $L^{-2j_x} \le O(|x|^{-2})$
and $L^{ -  (j_{x} - j_m)_+} \le O((m|x|)^{-1})$.
All this leads to
\begin{align}
\sum_{j = j_{x}}^\infty |R^{q_u}_j|
&\leq
L^{-2j_{x} - 2s (j_{x} - j_m)_+}
\sum_{j = j_{x}}^\infty O(\gbar_{j}) 4^{-(j - j_{x})}
\nnb
&\leq
|x|^{-2} (m|x|)^{-2s} O(\gbar_{j_{x}})
.
\label{e:vq-sum}
\end{align}
This gives the desired estimate \eqref{e:Rab-bound}.
\end{proof}

Thus, to prove Proposition~\ref{prop:R}, it
suffices to show that Theorem~\ref{thm:step-mr-fv} holds
with the $s$-dependent choice \refeq{elldef-zz}, for arbitrary $s>0$.
Constants in estimates will depend on $s$, and since we used
$s> \frac 12 (p+2)$ in the proof of Theorem~\ref{thm:mr}, such constants depend on $p$.


\section{Improved norm}
\label{sec:Rpf1}

The proof of Theorem~\ref{thm:step-mr-fv} is based on the observation that
it is possible to use the parameters \refeq{elldef-zz}
in the norm used in \cite{BS-rg-IE}, instead of the $s=0$ version used
previously.  In this section, we first
state improved covariance estimates, thereby indicating why it is possible
to improve the norm.
This leads to a discussion of simplified norm pairs beyond the mass
scale.  A lemma concerning the fluctuation-field regulator indicates why the
simplification is possible.
In the following, we use the notation appropriate for the spin field
$\varphi \in (\R^n)^\Lambda$ for $n \ge 1$; only notational modifications are needed for
$n=0$.

\subsection{Covariance bounds}
\label{sec:Cbds}


The estimate in \cite{ST-phi4} which yields the $s = 0$ case of \refeq{Rab-bound}
uses the norms defined in \cite{BS-rg-IE}.
One of these norms is the $\Phi_j(\ell_j)$ norm defined by
\begin{equation}
\lbeq{phinorm}
\|\varphi\|_{\Phi_j(\ell_j)}
=
\ell_j^{-1}
\sup_{x\in \Lambda}
\sup_{|\alpha|_1  \le p_\Phi}
L^{j|\alpha|_1}
|\nabla^{\alpha} \varphi_x|,
\end{equation}
which depends on the parameter $\ell_j$,
and on the maximal number of discrete derivatives $p_\Phi$
(fixed to be at least $4$ in \cite{BS-rg-IE}).
As in \refeq{elldef-zz}, we now define
\begin{align}
\label{e:elldef}
\ell_j &= \ell_0 L^{-j - s (j - j_m)_+}, \quad
\ell_{\sigma,j}
=
\ell_{j \wedge j_{x}}^{-1} 2^{(j - j_{x})_+} \ggen_j.
\end{align}
The analysis of \cite{BS-rg-IE,BS-rg-step} uses the norm parameters $\ell_j$ and $\ell_{\sigma,j}$ with $s = 0$.
To distinguish these from our
new choice \refeq{elldef} of $\ell_j$ and $\ell_{\sigma,j}$, we write
\begin{equation}
\label{e:ell-old}
    \ell_j^\old = \ell_0 L^{-j},
    \quad
    \ell_{\sigma,j}^\old  =
    (\ell_{j \wedge j_{x}}^{\rm old})^{-1}2^{(j - j_{x})_+}\ggen_j.
\end{equation}

In the more general terminology and notation of \cite{BS-rg-norm,BS-rg-IE},
we may regard a covariance $C_j$
in the decomposition \eqref{e:NCj}
as a test function depending on
two arguments $x,y$, and with this identification its $\Phi_j(\ell_j)$
norm is
\begin{equation}
    \label{e:Phinorm}
    \|C_j\|_{\Phi_{j}(\ell_j)}  =
    \ell_j^{-2}
    \sup_{x,y\in \Lambda}
    \;
    \sup_{|\alpha|_1 + |\beta|_1 \le p_\Phi}
    L^{(|\alpha|_1+  |\beta|_1)j}
    |\nabla_x^{\alpha} \nabla_y^{\beta} C_{j;x,y}|.
\end{equation}
The purpose of the $\Phi_j(\ell_j)$ norm is to measure the size of typical
fluctuation fields $\varphi$ with covariance $C_j$.
The parameter $\ell_j$ is chosen so that the norm of a typical field should
be $O(1)$, independent of $j$.

The following lemma justifies our choice of $\ell_j$
in \refeq{elldef}, by showing that the
bound \cite[\eqref{IE-e:CLbd}]{BS-rg-IE}, proved there only for the $s=0$ version
$\ell_j^\old$ of \refeq{ell-old},
remains true with the stronger
choice of norm parameter $\ell_j$ that permits arbitrary $s \ge 0$.
In its statement, the bounded sequence $\chicCov_j$ decays exponentially after the
mass scale and may be taken to be equal to
$2^{-(j-j_m)_+}$; its details are given
in \cite[Section~\ref{IE-sec:frp}]{BS-rg-IE} (where it is called $\chi_j$ rather
than $\chicCov_j$).

\begin{lemma}
[{Extension of \cite[\eqref{IE-e:CLbd}]{BS-rg-IE}}]
\label{lem:Cbd}
Given $\ellconst \in (0, 1]$, $\ell_0$ can be chosen large (depending on $L,\ellconst,s$)
so that
\begin{equation}
\lbeq{Cbd}
\|C_j\|_{\Phi_{j}(\ell_j)} \leq \min(\ellconst, \chicCov_j).
\end{equation}
\end{lemma}

The proof of Lemma~\ref{lem:Cbd} uses an estimate from
\cite[Proposition~\ref{pt-prop:Cdecomp}]{BBS-rg-pt}, which we repeat here as
the following proposition.

\begin{prop}[{Restatement of \cite[Proposition~\ref{pt-prop:Cdecomp}(a)]{BBS-rg-pt}}]
\label{prop:Cdecomp}
  Let $d >2$, $L\geq 2$, $j \ge 1$, $\bar m^2 >0$.
  For multi-indices $\alpha,\beta$ with
  $\ell^1$ norms $|\alpha|_1,|\beta|_1$ at most
  some fixed value $p$,
  and for any $k$, and for $m^2 \in [0,\bar m^2]$,
  \begin{equation}
    \label{e:scaling-estimate}
    |\nabla_x^\alpha \nabla_y^\beta C_{j;x,y}|
    \leq c(1+m^2L^{2(j-1)})^{-k}
    L^{-(j-1)(d-2 +|\alpha|_1+|\beta|_1)},
  \end{equation}
  where $c=c(p,k,\bar m^2)$ is independent of $m^2,j,L$.
  The same bound holds for $C_{N,N}$ if
  $m^2L^{2(N-1)} \ge \varepsilon$ for some $\varepsilon >0$,
  with $c$ depending on $\varepsilon$ but independent of $N$.
\end{prop}

\begin{proof}[Proof of Lemma~\ref{lem:Cbd}]
For $d=4$, insertion of \refeq{scaling-estimate} into \refeq{Phinorm} gives
\begin{equation}
    \label{e:Phinorm2}
    \|C_j\|_{\Phi_{j}(\ell_j)}
    \le
    c
    L^{p_\Phi}
    \ell_j^{-2}(1+m^2L^{2(j-1)})^{-k}
    L^{-2(j-1)}.
\end{equation}
With $s=0$ in \eqref{e:elldef}, \refeq{Phinorm2} gives
$\|C_j\|_{\Phi_{j}(\ell_j)} \le c_L \ell_0^{-2} (1+m^2L^{2(j-1)})^{-k}$
for an $L$-dependent constant $c_L$ (whose value may now change from line to line).
The estimate \cite[\eqref{IE-e:CLbd}]{BS-rg-IE}
is wasteful in that it does not make any use of the factor
$(1+m^2L^{2(j-1)})^{-k}$ in \refeq{Phinorm2} beyond extraction of the factor $\chicCov_j$.
To improve this, we now allow arbitrary $s$, and fix the arbitrary parameter $k$ to be $k=s+1$
in \refeq{Phinorm2} so that
\begin{equation} \label{e:mass-decay}
(1 + m^2 L^{2j})^{-k} \le c_L L^{-2(s+1)(j - j_m)_+}.
\end{equation}
We insert \refeq{mass-decay} and the definition $\ell_j=\ell_0 L^{-j-s(j-j_m)_+}$ from
\refeq{elldef} into
\eqref{e:Phinorm2}, to conclude that there exists $c_0 = c_0(s, L)$ such that
\begin{equation}
    \|C_j\|_{\Phi_{j}(\ell_j)} \leq c_0 \ell_0^{-2} L^{-2(j - j_m)_+}
    .
\end{equation}
By definition of $\chicCov_j$ (see \cite[Section~\ref{IE-sec:frp}]{BS-rg-IE}),
$L^{-2(j - j_m)_+}$ is bounded by a multiple of $\chicCov_j$.
It thus suffices to choose $\ell_0$ large enough that
$\ell_0^2 \ge c_0 \ellconst^{-1}$.
\end{proof}

\subsection{New choice of norm beyond the mass scale}
\label{sec:newnorm}

As in \cite[(\ref{IE-e:PhiXdef})]{BS-rg-IE}, we use the localised version
of \eqref{e:phinorm}, defined for subsets
$X \subset \Lambda$  by
\begin{align}
\label{e:PhiXdef}
    \|\varphi\|_{\Phi_j(X)}
    &=
    \inf \{ \|\varphi -f\|_{\Phi_j} :
    \text{$f \in \C^\Lambda$ such that $f_{x} = 0$
    $\forall x\in X$}\}.
\end{align}
A \emph{small set} is defined to be a
connected polymer $X \in \Pcal_j$ consisting of at most $2^d$ blocks
(the specific number
$2^d$ plays no direct role here),
and $\Scal_j \subset \Pcal_j$ denotes the set of small sets.
The \emph{small set neighbourhood} of $X \subset \Lambda $ is
the enlargement of $X$ defined by
$
    X^{\Box}
=
    \bigcup_{Y\in \Scal_{j}:X\cap Y \not =\varnothing } Y$.

Given $X \subset \Lambda$ and $\varphi \in (\R^n)^{\Lambda}$,
we recall from \cite[\eqref{IE-e:GPhidef}]{BS-rg-IE}
that the
\emph{fluctuation-field regulator} $G_j$
is defined by
\begin{align}
\label{e:GPhidef}
    G_j(X,\varphi)
    =
    \prod_{x \in X} \exp
    \left(|B_{x}|^{-1}\|\varphi\|_{\Phi_j (B_{x}^\Box,\ell_j )}^2 \right)
    ,
\end{align}
where $B_{x}\in \Bcal_j$ is the unique block that contains $x$,
and hence $|B_x| = L^{dj}$.
The \emph{large-field regulator} is defined in \cite[\eqref{IE-e:9Gdef}]{BS-rg-IE} by
\begin{align}
\label{e:9Gdef}
    \tilde G_j  (X,\varphi)
    =
    \prod_{x \in X}
    \exp \left(
    \frac 12 |B_{x}|^{-1}\|\varphi\|_{\tilde\Phi_j (B_{x}^\Box,\ell_j)}^2
    \right)
    .
\end{align}
The $\tilde\Phi_j$ norm appearing on the right-hand side of \refeq{9Gdef} is
similar to the $\Phi_j$ norm, with the important difference that it is insensitive to
shifts by linear test functions; see \cite[\eqref{IE-e:Phitilnorm}]{BS-rg-IE} for the
precise definition.
The two regulators serve as weights in the \emph{regulator norms} of
\cite[Definition~\ref{IE-def:Gnorms}]{BS-rg-IE}.
The regulator norms are defined,  with $\gamma \in (0,1]$ and
for $F$ in the space $\Ncal(X^\Box)$ of functionals
of the field (see \cite[\eqref{norm-e:NXdef}]{BS-rg-norm}), by
\begin{align}
\label{e:Gnormdef1}
    \| F\|_{G_j(\ell_j)}
    &=
    \sup_{\varphi \in (\R^n)^\Lambda}
    \frac{\|F\|_{T_{\varphi,j}(\ell_j)}}{G_{j}(X,\varphi)}
    ,
\\
\label{e:Gnormdef2}
    \|F\|_{\tilde G_j^{\Gtilp}(h_j)}
    &=
    \sup_{\varphi \in (\R^n)^\Lambda}
    \frac{\|F \|_{T_{\varphi,j}(h_j)}}{\tilde{G}^{\Gtilp}_{j}(X,\varphi)}
    .
\end{align}
The parameter $\ell_j$ that appears in the regulators \refeq{GPhidef}--\refeq{9Gdef} and
in the numerator of \refeq{Gnormdef1} was taken to be $\ell_j^\old$ in \cite{BS-rg-IE},
but now we use $\ell_j$ instead. As in \cite{BS-rg-IE},
the parameter $h_j$ and its observable counterpart $h_{\sigma,j}$ are given by
\begin{align}
\label{e:h}
    h_{j} &= k_0 \ggen_j^{-1/4}L^{-j},
    \quad
    h_{\sigma,j}  = (\ell_{j \wedge j_{x}}^{\rm old})^{-1}
    2^{(j - j_{x})_+}\ggen_j^{1/4}.
\end{align}

In \cite{BS-rg-IE}, estimates on $\|\cdot\|_{j+1}$ are given in terms of
$\|\cdot\|_j$, where the pair $(\|\cdot\|_j, \|\cdot\|_{j+1})$ refers to
either of the norm pairs
\begin{equation}
\label{e:np1}
    \|F\|_j = \|F\|_{G_j(\ell_j^\old)}
    \quad \text{and} \quad
    \|F\|_{j+1} = \|F\|_{T_{0,j+1}(\ell_{j+1}^\old)},
\end{equation}
or
\begin{equation}
\label{e:np2}
    \|F\|_j = \|F\|_{\tilde{G}_j(h_j)}
    \quad \text{and} \quad
    \|F\|_{j+1} = \|F\|_{\tilde{G}_{j+1}^{\Gtilp}(h_{j+1})}.
\end{equation}
We will show that, \emph{above the mass scale} $j_m$ (see \eqref{e:jmdef}), the results of \cite{BS-rg-IE} hold  with
both norm pairs in \eqref{e:np1} and \eqref{e:np2} replaced by the single new norm pair
\begin{equation}
\label{e:npmass}
    \|F\|_j = \|F\|_{G_j(\ell_j)}
    \quad \text{and} \quad
    \|F\|_{j+1} = \|F\|_{G_{j+1}(\ell_{j+1})},
\end{equation}
with the improved $\ell_j$ of \eqref{e:elldef} with $s>0$ fixed as large as desired.

The space $\Ncal$ containing the functionals $F$ appearing above requires control on
up to $p_\Ncal$ derivatives of $F$ with respect to the field $\varphi$,
where $p_\Ncal$ is a parameter of the $T_\varphi$-norm.
In the proof of Proposition~\ref{prop:cl} below,
we must choose $p_\Ncal$ to be large depending on $p$,
in order to analyse the correlation length
of order $p$.  The renormalisation group analysis is predicated on fixed (but arbitrary)
$p_\Ncal$, so it can proceed with this modification.  However,
we do not prove that constants are uniform in $p_\Ncal$,
and in particular we do not prove that the required smallness of $g$ in
Theorem~\ref{thm:mr} is uniform in the choice of $p_\Ncal$.
Thus we do not have a result for \emph{all} $p>0$ for any fixed $g$.

The use of two norm pairs adds intricacy to \cite{BS-rg-IE,BS-rg-step}.
The pair \refeq{np1} is insufficient, on its own, because the scale-$(j+1)$ norm
is the $T_0$ semi-norm which controls only small fields, and an estimate in this norm
does not imply an estimate for the $G_{j+1}$ norm.  The norm pair \refeq{np2} is
used to supplement the norm pair \refeq{np1}, and estimates in both of the scale-$(j+1)$
norms can be combined to provide an estimate for the $G_{j+1}$ norm.  This then
sets the stage for the next renormalisation group step.  Above the mass scale,
the use of \refeq{npmass} now bypasses many issues.  For example, for $j>j_m$
 the $\Wcal_j$ norm of \cite[\eqref{step-e:9Kcalnorm}]{BS-rg-step} is replaced
 simply by the $\Fcal_j(G)$ norm, and there is no need for the $\Ycal_j$ norm of
\cite[\eqref{step-e:Ycaldef}]{BS-rg-step} nor for \cite[Lemma~\ref{step-lem:KKK}]{BS-rg-step}.

The need for both norm pairs \eqref{e:np1}--\eqref{e:np2} is discussed in
\cite[Section~\ref{IE-sec:lfp}]{BS-rg-IE} and is related to the
so-called \emph{large-field problem}. Roughly speaking, the
norm pair \refeq{np2} is used to take advantage of the quartic term in the interaction to
suppress the effects of large values of the fields. This approach
relies on the fact that the interaction polynomial is dominated by the
quartic term in the $h$-norm, as expressed by
\cite[\eqref{IE-e:tau2dom}]{BS-rg-IE}, together with the lower bound
\cite[\eqref{IE-e:epVbark0}]{BS-rg-IE} on the quartic term.
However, above the mass scale, large fields are naturally suppressed
by the rapid decay of the covariance.
This idea is captured in Lemma~\ref{lem:mart} below, which replaces
\cite[Lemma~\ref{IE-lem:mart}]{BS-rg-IE} above the mass scale.
The regulators in its statement are defined by \refeq{GPhidef} with the $s$-dependent
$\ell_j$ of \refeq{elldef}.

\begin{lemma}[{Replacement for \cite[Lemma~\ref{IE-lem:mart}]{BS-rg-IE}}]
\label{lem:mart}
Let $X \subset \Lambda$ and assume that $s > 1$.
For any $q >0$, if $L$ is sufficiently large depending on $q$, then for $j_m \leq j < N$,
\begin{equation}
\label{e:mart}
    G_{j}(X, \varphi)^{q}
    \le
    G_{j+1}(X, \varphi).
\end{equation}
\end{lemma}
\begin{proof}
By \eqref{e:GPhidef}, it suffices to show that, for any scale-$j$ block $B_j$ and any scale-$(j+1)$ block $B_{j+1}$ containing $B_j$,
\begin{equation}
q \|\varphi\|^2_{\Phi_j (B_j^\Box,\ell_j )}
\leq
L^{-4} \|\varphi\|^2_{\Phi_{j+1} (B_{j+1}^\Box,\ell_{j+1})}.
\end{equation}
In fact, since $\|\varphi\|_{\Phi_j (B_j^\Box,\ell_j )}
\leq \|\varphi\|_{\Phi_j (B_{j+1}^\Box,\ell_j )}$ by definition,
it suffices to prove the above bound with $B_j$ replaced by $B_{j+1}$ on the left-hand side.
According to the definition of the norm in \eqref{e:PhiXdef},
to show this it suffices to prove that
\begin{equation}
\lbeq{martwant}
q \|\varphi\|_{\Phi_j(\ell_j)}^2 \leq L^{-4} \|\varphi\|_{\Phi_{j+1}(\ell_{j+1})}^2
\end{equation}
(then we replace $\varphi$ by $\varphi -f$ in the above and take the infimum).

By definition,
\begin{equation}
\|\varphi\|_{\Phi_j(\ell_j)}
\le
\ell_j^{-1} \ell_{j+1}
\sup_{x\in \Lambda} \sup_{|\alpha| \leq p_\Phi}
\ell_{j+1}^{-1}
L^{(j+1) |\alpha|}
|\nabla^\alpha \varphi_x|,
\end{equation}
with the inequality due to replacement of $L^{j |\alpha|}$ on the left-hand
side by $L^{(j+1) |\alpha|}$ on the right-hand side.
Since $\ell_j^{-1} \ell_{j+1} = L^{-1 - s \1_{j \geq j_m}}$,
\begin{equation}
\|\varphi\|_{\Phi_j(\ell_j)} \leq L^{-1 - s \1_{j \geq j_m}} \|\varphi\|_{\Phi_{j+1}(\ell_{j+1})}.
\end{equation}
Thus,
\begin{equation}
q \|\varphi\|_{\Phi_j(\ell_j)}^2
\leq q L^{-4} L^{2 - 2s \1_{j \geq j_m}} \|\varphi\|^2_{\Phi_{j+1}(\ell_{j+1})},
\end{equation}
and then \refeq{martwant}
follows once $L$ is large enough that $q L^{2 - 2s} \leq 1$.
\end{proof}

\begin{rk}
The elimination of the $h$-norm after the mass scale is more than a convenience.
It becomes a necessity when we improve the $\ell$-norm.
Briefly, the reason is as follows. In the proof of
\cite[Lemma~\ref{step-lem:KKK}]{BS-rg-step}, the ratio
$\ell_{\sigma}/h_{\sigma}$
must be bounded. For this, we would need
to increase $h_{\sigma}$
beyond the mass scale  (since $\ell_{\sigma}$ has been increased).
This forces a compensating decrease in $h$
beyond $j_m$, to keep the product $hh_{\sigma}$ bounded for stability (as in
Section~\ref{sec:stability1} below).
But if we do this, we lose the lower bound required on $\epsilon_{g\tau^2}$
required for stability in the $h$-norm (see \cite[\eqref{IE-e:epVbardefz-app}]{BS-rg-IE}).
\end{rk}

\section{Proof of Theorem~\ref{thm:step-mr-fv}}
\label{sec:Rpf2}

In this section, we show that
Theorem~\ref{thm:step-mr-fv}
holds,
thereby completing the proof of Proposition~\ref{prop:R}.
The key steps in
the proof of the $s = 0$ case of Theorem~\ref{thm:step-mr-fv}
are contained in \cite{BS-rg-IE,BS-rg-step}.
Our main objective in this section is to show that the results in \cite{BS-rg-IE,BS-rg-step}
continue to hold with the new norm parameters $\ell_j,\ell_{\sigma,j}$.
To this end, we may and do use the fact that the estimates of \cite{BS-rg-IE} have already been
established with the old norm parameters.

In the following,
we indicate the changes in the analysis of
\cite{BS-rg-IE,BS-rg-step} that arise due to the new choice of norm parameters \refeq{elldef}
beyond the mass scale, and due to the reduction from two norm pairs to one.
This requires repeated reference to previous papers.

\subsection{Norm parameter ratios}
\label{sec:norm-parameter-ratios}

The analysis of \cite{BS-rg-IE} assumes that the norm parameters $\h_j,\h_{\sigma,j}$,
for $\h = \ell$ or $\h = h$,
satisfy the estimates \cite[\eqref{IE-e:h-assumptions}]{BS-rg-IE}; these assert that
\begin{align}
\label{e:h-assumptions-IE}
    \h_j \ge \ell_{j},
    \quad\quad
    \frac{\h_{j+1}}{\h_j}
    &\le 2 L^{-1},
    \quad\quad
    \frac{\h_{\sigma,j+1}}{\h_{\sigma,j}}
    \le
    {\rm const}\,
    \begin{cases}
     L  & (j < j_{x})
     \\
     1 & (j \ge j_{x}).
     \end{cases}
\end{align}
We do not change $\h_j$ or $\h_{\sigma,j}$ for $j$ below the mass scale, so there
can be no difficulty until above the mass scale.  Above the mass scale, the parameters
$h_j,h_{\sigma,j}$ are eliminated, and requirements involving them become vacuous.
Thus, for \refeq{h-assumptions-IE}, we need only verify the second and third
inequalities for the case $\h=\ell$.
By definition,
\begin{equation}\label{e:h-assumptions}
\frac{\ell_{j+1}}{\ell_j} = L^{-(1 + s \1_{j \geq j_m})},
\qquad
\frac{\ell_{\sigma, j+1}}{\ell_{\sigma,j}} = \frac{\ggen_{j+1}}{\ggen_j}
\times
\begin{cases} L^{1 + s \1_{j \geq j_m}} & (j < j_x) \\ 2 & (j \geq j_x). \end{cases}
\end{equation}
According to \cite[\eqref{IE-e:gbarmono}]{BS-rg-IE},
$\frac 12 \ggen_{j+1} \le \ggen_j \le 2 \ggen_{j+1}$.
Thus, the second estimate of \eqref{e:h-assumptions-IE}
is satisfied (the ratio being improved when $j\ge j_m$),
while the third is \emph{not} when $s >0$ and $j_m < j_x$.
This potentially dangerous third
estimate in \refeq{h-assumptions-IE} is used to prove the scale monotonicity lemma
\cite[Lemma \ref{IE-lem:Imono}]{BS-rg-IE}, as well
as the crucial contraction.
We discuss
\cite[Lemma \ref{IE-lem:Imono}]{BS-rg-IE} next, and return to the crucial contraction
in Section~\ref{sec:cc} below.

\paragraph{\cite[Lemma \ref{IE-lem:Imono}]{BS-rg-IE}}
There is actually no problem with the scale monotonicity lemma.
Indeed, for the case $\alpha =ab$ of the proof of
\cite[Lemma \ref{IE-lem:Imono}]{BS-rg-IE}, the hypothesis that
$\pi_{0x}F=0$ for $j<j_{x}$ ensures that this case only relies on the dangerous
estimate for $j \ge j_x$ where the danger is absent in \refeq{h-assumptions}.
For the cases $\alpha=a$ and $\alpha =b$ of the proof of
\cite[Lemma \ref{IE-lem:Imono}]{BS-rg-IE}, what is
important is the inequality
$\ell_{\sigma,j+1}\ell_{j+1} \le {\rm const}\, \ell_{\sigma,j}\ell_{j}$, which
continues to hold with
\refeq{elldef} for all scales $j$, both above and below the mass scale, since
the products in this inequality are the same for the new and the old choices of $\ell$.
So \cite[Lemma \ref{IE-lem:Imono}]{BS-rg-IE} continues to hold with the choice
\refeq{elldef}.
In addition,
\begin{equation}
\label{e:norm-change}
\|F\|_{T_\varphi(\ell_j)} \leq  \|F\|_{T_\varphi(\ell^\old_j)}.
\end{equation}
This strengthened special case of the first inequality of
\cite[\eqref{IE-e:scale-change}]{BS-rg-IE} (strengthened due to the constant
$1$ on the right-hand side of \refeq{norm-change} compared to
the generic constant in \cite[\eqref{IE-e:scale-change}]{BS-rg-IE})
can be seen from an examination of the proof of the $\alpha =a,b$ case of
\cite[Lemma \ref{IE-lem:Imono}]{BS-rg-IE}, together with the observation that
$\ell_{\sigma,j}\ell_{j} = \ell_{\sigma,j}^{\rm old} \ell_j^{\rm old}$ by definition.

\subsection{Stability domains}
\label{sec:stability1}

The stability domain $\DV_j$ is defined in \cite[\eqref{IE-e:DV1-bis}]{BS-rg-IE}.
We modify $\DV_j$ only for the coupling constant $q$,
by replacing $r_q$ in
\cite[\eqref{IE-e:h-coupling-def-1-bis}]{BS-rg-IE} by
\begin{equation}
\lbeq{newrq}
L^{2j_x + 2 s (j_x - j_m)_+} 2^{2(j-j_x)} r_{q,j} =
\begin{cases}
  0 & j < j_x \\
  C_{\DV}  & j \ge j_x.
\end{cases}
\end{equation}

\paragraph{\cite[Proposition~\ref{IE-prop:monobd}]{BS-rg-IE}}
With \refeq{newrq}, \cite[Proposition~\ref{IE-prop:monobd}]{BS-rg-IE}
as it pertains to $\h=\ell$ (omitting all reference to $\h=h$) continues to hold beyond
the mass scale by the same proof.
In particular, with the smaller choice for the domain of $q$,
\cite[\eqref{IE-e:qhsig}]{BS-rg-IE} holds with the larger $s$-dependent $\ell_{\sigma,j}$.

\medskip
Note that we do not need to change the domain of $\lambda$.
This is because the bound \cite[\eqref{IE-e:hsigh}]{BS-rg-IE}
continues to hold with the new norm parameters. Indeed, while $\ell_j$
and $\ell_{\sigma,j}$ have been modified, their product $\ell_j \ell_{\sigma,j}$
has not.
This guarantees that the $T_0$ semi-norm
$\|\sigma\bar\varphi_a\|_{T_0} = \ell_\sigma \ell$ remains identical to what it was
with the old norm parameters, and therefore there is no new stability requirement
arising from this.

The choice \refeq{newrq} places a more stringent requirement on the domain
than does the $s=0$ version.  To see that this requirement is actually met
by the renormalisation group flow,
we note a minor improvement to the proof of \cite[Lemma~\ref{step-lem:K7a}(ii)]{BS-rg-step},
where the bound $|\delta q| \leq c L^{-2j}$ is used to show that $v(X)$
(defined there) satisfies
\begin{equation}
\lbeq{vXbd}
\|v(X)\| \leq c L^{-2j} (\ell_{\sigma,j}^\old)^2 \leq c'.
\end{equation}
Here the factor $L^{-2j}$ arises
as a bound on the covariance $C_{j+1;00}$ in the perturbative flow
\cite[\eqref{pt-e:qpt2}]{BS-rg-IE} of $q$ and it can therefore be improved to $L^{-2j-2s(j - j_m)_+}$
by Lemma~\ref{lem:Cbd}.
Thus also with $\ell^\old, \ell_{\sigma}^{\rm old}$ replaced by $\ell,\ell_{\sigma}$,
the required bound $\|v(X)\| \leq c'$ remains valid.

\subsection{Extension of stability analysis}
\label{sec:stability2}

In this and the next section,
we verify that the results of \cite[Section~\ref{IE-sec:IE}]{BS-rg-IE}
remain valid with $\ell^\old$ replaced by $\ell$.
In this section, we deal with the results whose proofs need only minor
modification.

First, we note that the supporting results of \cite[Section~\ref{IE-sec:W}]{BS-rg-IE} hold with the new norms.
Indeed,
it is immediate from \eqref{e:norm-change} that analogues of
\cite[Proposition~\ref{IE-prop:Wnorms}]{BS-rg-IE} and
\cite[Lemmas~\ref{IE-lem:epdV}, \ref{IE-lem:W-logwish}--\ref{IE-lem:Wbil}]{BS-rg-IE} hold with the new $\ell_j$.
Moreover, \cite[Lemma~\ref{IE-lem:Fpibd-bis}]{BS-rg-IE} and \cite[Proposition~\ref{IE-prop:Wbounds}]{BS-rg-IE} hold
for general values of the parameters $\h_j$ (which are implicit in the $T_{0,j}$-norm).
We discuss \cite[Proposition~\ref{IE-prop:1-LTdefXY}]{BS-rg-IE}
in Section~\ref{sec:cc} below,
and the remaining results of \cite[Section~\ref{IE-sec:W}]{BS-rg-IE} do not make use of norms.

\paragraph{\cite[Proposition~\ref{IE-prop:Iupper}]{BS-rg-IE}}
With $\h = \ell$, \cite[\eqref{IE-e:Iupper-a}]{BS-rg-IE} continues to hold with the same proof;
in fact the proof does not depend on the explicit choice of $\h$.
We do not need \cite[\eqref{IE-e:Iupper-b}]{BS-rg-IE} as it is only applied with $\h = h$.

\paragraph{\cite[Proposition~\ref{IE-prop:Istab}]{BS-rg-IE}}
The only change to the proof is for the case $j_* = j + 1$.
To get \cite[\eqref{IE-e:IF}]{BS-rg-IE},
we proceed as previously in the case $\h=h$ but applying Lemma~\ref{lem:mart}
rather than \cite[Lemma~\ref{IE-lem:mart}]{BS-rg-IE} following \cite[(5.22)]{BS-rg-IE}.
In the same way, we get \cite[\eqref{IE-e:Iass}]{BS-rg-IE} and the remaining parts of
the proposition follow without changes to the proof.

\paragraph{\cite[Proposition~\ref{IE-prop:Ianalytic1:5}]{BS-rg-IE}}
Again the only required change in the proof is the use of
Lemma~\ref{lem:mart} in the case $j_* = j + 1$,
for which as previously we use Lemma~\ref{lem:mart} instead of \cite[Lemma~\ref{IE-lem:mart}]{BS-rg-IE}.

\paragraph{\cite[Proposition~\ref{IE-prop:JCK-app-1}]{BS-rg-IE}}
No changes need to be made to the proof.
In fact, it is necessary \emph{not} to use the $\h = \ell$ case
of the estimate \cite[(5.32)]{BS-rg-IE}. Instead, the
$\h = \ell^\old$ case of this estimate should be used for $g_Q$.
This is possible since the renormalisation
group map, and in particular the coupling constants, are independent of the choice of norm.

\paragraph{\cite[Proposition~\ref{IE-prop:hldg}]{BS-rg-IE}}
Using \refeq{norm-change}, we see that the proof
continues to hold
above the mass scale.
The only change to the proof is that in the application of
\cite[Proposition~\ref{IE-prop:Istab}]{BS-rg-IE}, $j$ should be replaced by
$j + 1$ in \cite[\eqref{IE-e:IF}]{BS-rg-IE} with $j_* = j + 1$ (corresponding
to the $G_{j+1}$ norm). This yields \cite[\eqref{IE-e:2Lprimeh1}]{BS-rg-IE}
with a $G_{j+1}$ norm on the left-hand side.

\paragraph{\cite[Proposition~\ref{IE-prop:h}]{BS-rg-IE}}
A version of \cite[Lemma~\ref{IE-lem:dIipV}]{BS-rg-IE} with the new $\ell$
continues to hold. This lemma makes use of $\hat\ell$,
which superficially depends on the choice of $\ell$ in its definition
\cite[\eqref{IE-e:ellhatdef}]{BS-rg-IE}. However, brief scrutiny of
\cite[\eqref{IE-e:ellhatdef}]{BS-rg-IE} reveals that
the apparent dependence on $\ell$ actually cancels and there is in fact no dependence.
Similarly,
\cite[Lemma~\ref{IE-lem:epdV}]{BS-rg-IE} continues to hold without
any changes to its proof.
The proof of \cite[Proposition~\ref{IE-prop:h}]{BS-rg-IE} then applies without change.

\paragraph{\cite[Proposition~\ref{IE-prop:ip}]{BS-rg-IE}}
With the new choice of $\ell$ (and $\Gcal = G$),
\cite[Lemma~\ref{IE-lem:dIip}]{BS-rg-IE} continues to hold with no changes to its proof.
Thus, by \cite[\eqref{IE-e:scale-change}]{BS-rg-IE}
and \cite[Lemma~\ref{IE-lem:dIip}]{BS-rg-IE},
\begin{align}
  \label{e:integration-property-pf}
     &\|\Ex_{j+1} \delta I^X \theta F(Y) \|_{T_{\varphi,j+1}(\ell_{j+1})}
     \nnb
     &\quad \leq
     \|\Ex_{j+1} \delta I^X \theta F(Y) \|_{T_{\varphi,j}(\ell_{j})} \nnb
     &\quad \leq
     \Econst^{|X|_j+|Y|_j}
     (C_{\delta V} \epdV)^{|X|_j}
     \| F(Y) \|_{G_j(\ell_{j})}
    G_j(X\cup Y,\varphi)^5
    .
\end{align}
By Lemma~\ref{lem:mart}, $G_j(X\cup Y,\varphi)^5 \le G_{j+1}(X\cup Y,\varphi)$.
Now we divide both sides by
$G_{j+1}(X \cup Y, \varphi)$ and take the supremum over $\varphi$ to complete the proof.

\subsection{Extension of the crucial contraction}
\label{sec:cc}

The proof of the ``crucial contraction''
\cite[Proposition \ref{IE-prop:cl}]{BS-rg-IE}
makes use of the  third estimate in
\eqref{e:h-assumptions-IE}, which is now violated above the mass scale
due to our new choice of $\ell_j$.
On the other hand, the second estimate of \eqref{e:h-assumptions-IE} is
improved by the new choice and compensates for the degraded third estimate,
as we explain in this section.

Below the mass scale, we continue to use the crucial contraction as stated in
\cite[Proposition \ref{IE-prop:cl}]{BS-rg-IE} in terms of two norm pairs.
Next, we state a version of the crucial contraction for use above the mass
scale using the new norm pair \eqref{e:npmass}.
The statement uses the notation of \cite{BS-rg-IE} (which we do not redefine here),
with the exception that now we have replaced $a$ by $0$, $b$ by $x$, and $j_{ab}$ by $j_x$
for consistency with our present notation.
Throughout this section, we
sometimes write the dimension as $d$ for emphasis, although we only consider $d=4$.

\begin{prop}[{Improvement of \cite[Proposition \ref{IE-prop:cl}]{BS-rg-IE}}]
\label{prop:cl} Let $j_m \leq j<N$ and $V\in \DV_j$.  Let $X \in \Scal_j$ and
$U = \overline X$.  Let $F(X) \in \Ncal(X^\Box)$ be such that
$\pi_\alpha F(X) =0$ when $X(\alpha)=\varnothing$, and such that
$\pi_{0x}F(X)=0$ unless $j \ge j_x$.
{There is a constant $C$ (independent of $L$) such that}
\begin{align}
    \label{e:contraction3z-new}
    \|\Ipttil^{U\setminus X} \Ex_{C_{j+1}} \theta F (X) \|_{{G_{j+1}(\ell_{j+1})}}
    &
    \leq C \Big(
    ( L^{-d -1} +  L^{-1}\1_{X \cap \{0,x\} \not = \varnothing} )
    \kappa_F
    + \kappa_{\LT F}
    \Big)
    ,
\end{align}
with $\kappa_F=\|F (X)\|_{{G_{j}(\ell_{j})}}$ and
$\kappa_{\LT F} =\|\Ipttil^X \LT_X \Ipttil^{-X} F(X) \|_{{G_{j}(\ell_{j})}}$.
\end{prop}

An ingredient in the proof of Proposition~\ref{prop:cl} is
\cite[Lemma~\ref{loc-lem:phij}]{BS-rg-loc}, which is the $s=0$ version of
the following lemma.  For simplicity, we state only the conclusion of the lemma,
and the notation and hypotheses are those in
\cite[Lemma~\ref{loc-lem:phij}]{BS-rg-loc}, except now we use the
$s$-dependent norm parameters
$\h_j=\ell_j$ of \refeq{elldef} ($h_j$ is not needed above the mass scale, and
the $s=0$ case applies below the mass scale).

\begin{lemma}[{Improvement of \cite[Lemma~\ref{loc-lem:phij}]{BS-rg-loc}}]
\label{lem:phiij-improved}
With the same hypotheses and notation as in \cite[Lemma~\ref{loc-lem:phij}]{BS-rg-loc},
\begin{equation}
\lbeq{gbd-improved}
\|g\|_{\tilde{\Phi} (X)}
\leq
\bar C_3
L^{- (1 + s \1_{j \geq j_m}) d_+'}  \|g\|_{\tilde{\Phi}' (X_+)}.
\end{equation}
\end{lemma}

\begin{proof}
The proof of \cite[Lemma~\ref{loc-lem:phij}]{BS-rg-loc}
is based on the assumption  $\ell_{j+1}/\ell_j \leq  cL^{-1}$
(we take $[\varphi_i]=1$; the parameters $\ell_{\sigma,j}$ are not used).
For our new values of $\ell$, the stronger assumption
$\ell_{j+1}/\ell_j \leq L^{-1 - s \1_{j \geq j_m}}$ holds.
The unique change to the proof occurs in the transition from
\cite[\eqref{loc-e:gTay1}]{BS-rg-loc} to
\cite[\eqref{loc-e:rhognew}]{BS-rg-loc}, where the ratio
$\ell_{j+1}/\ell_j$ is used.
With the new ratio, \cite[\eqref{loc-e:rhognew}]{BS-rg-loc} becomes
\begin{align}
\label{e:rhognew-improved}
    \|r\|_{\Phi (X)}
    & \le
    \sup_{z \in {\mathbf X}_+}
    (cK\ell'^{-1})^z
    \sup_{|\beta|_\infty \le p_\Phi}
    L^{-(p(z) + p(z) s \1_{{j\geq j_m}} +|\beta|_1)}
    | \nabla_{R'}^\beta  r_{z}  |.
\end{align}
Here $r = h - \Tay_a h$, where $h$ is an arbitrary test function and $a$ is the largest
point which is lexicographically no larger than any point in $X$.
The test function $h$ depends on sequences of points $(x_1, \dots, x_p)$,
and $\Tay_a h$ is a discrete version of Taylor's approximation which approximates $h$ by a
discrete Taylor polynomial localised at point $a$ in each argument (see \cite{BS-rg-loc} for details).
By definition, for the empty sequence $\varnothing$, $(\Tay_a h)_\varnothing = h_\varnothing$,
and thus
$r_\varnothing = 0$.

It follows that we can take $p(z) \geq 1$ in the supremum over $z \in \mathbf{X}_+$ in \eqref{e:rhognew-improved}. Thus,
\begin{align}
    \|r\|_{\Phi (X)}
    & \le
    L^{-s \1_{{j\geq j_m}}}
    \sup_{z \in {\mathbf X}_+}
    (cK\ell'^{-1})^z
    \sup_{|\beta|_\infty \le p_\Phi}
    L^{-(p(z) +|\beta|_1)}
    | \nabla_{R'}^\beta  r_{z}  |.
\end{align}
The quantity
\begin{align}
\label{e:rhognew}
    \sup_{z \in {\mathbf X}_+}
    (cK\ell'^{-1})^z
    \sup_{|\beta|_\infty \le p_\Phi}
    L^{-(p(z) +|\beta|_1)}
    | \nabla_{R'}^\beta  r_{z}  |
\end{align}
is identical to the right-hand side of \cite[\eqref{loc-e:rhognew}]{BS-rg-loc} when $[\varphi_i] = 1$. In \cite{BS-rg-loc}, it is shown that this quantity can be bounded by a constant times
\begin{align}
    L^{-d_{+}'}
    \|h\|_{\Phi'( X_+)}.
\end{align}
Thus,
\begin{equation}
    \|r\|_{\Phi (X)}
    \leq \bar C_3
    L^{-s \1_{{j\geq j_m}}}
    L^{-d_{+}'}
    \|h\|_{\Phi'( X_+)}.
\end{equation}
With this improvement to \cite[\eqref{loc-e:rhognew}]{BS-rg-loc} in the proof of
\cite[Lemma~\ref{loc-lem:phij}]{BS-rg-loc}, the conclusion of
\cite[Lemma~\ref{loc-lem:phij}]{BS-rg-loc} is improved to \refeq{gbd-improved}.
\end{proof}

Roughly speaking, the $L$-dependent factor in \eqref{e:gbd-improved} implements the dimensional gain
for irrelevant directions in a renormalisation group step, when passing from one scale to the next.
In other words, we may regard the dimension of the field as improving from $1$ below the
mass scale to $1+s$ above the mass scale.
The $s=0$ version of Lemma~\ref{lem:phiij-improved} is adapted to the scaling at the critical point, where $m^2=0$.
In the noncritical case $m^2>0$, the dimensional gain improves greatly for $j>j_m$,
as apparent from \eqref{e:scaling-estimate}, and is
captured more accurately by the general-$s$ version of \eqref{e:gbd-improved}.

As a consequence of the former improvement we have the following two further improvements.
From now on, we always assume $\h=\ell$ and $j>j_m$, as this is the only case relevant for
the improvement of \cite[Proposition~\ref{IE-prop:cl}]{BS-rg-IE}.

\paragraph{\cite[Proposition~\ref{loc-prop:1-LTdefXY}]{BS-rg-loc}}
The improvement in Lemma~\ref{lem:phiij-improved} propagates to
\cite[Proposition~\ref{loc-prop:1-LTdefXY}]{BS-rg-loc}, which now holds
as stated except with
$\gamma_{\alpha,\beta}$
improved to
\begin{equation}
\label{e:cgamobs}
    \gamma_{\alpha,\beta}
        =
    \left(
    L^{-(d_\alpha' + s \1_{j \geq j_m})} +  L^{-(A+1)}
    \right)
    \left( \frac{\ell_{\sigma,j+1}}{\ell_{\sigma,j}} \right)^{|\alpha \cup \beta|}
    .
\end{equation}
The right-hand side can be estimated as follows.
By \eqref{e:h-assumptions},
\begin{equation}
\frac{\ell_{\sigma,j+1}}{\ell_{\sigma,j}} \leq
4
  \begin{cases}
  L^{1 + s \1_{{j \geq j_m}}} & j < j_x \\
  1 & j \geq j_x,
  \end{cases}
\end{equation}
and hence
\begin{equation}
\label{e:cgamobs-L}
    \gamma_{\alpha,\beta}
    \le C''
    \left(
    L^{-(d_\alpha' + s \1_{j \geq j_m})} +  L^{-(A+1)}
    \right)
    \times
    \begin{cases}
    L^{(1 + s \1_{{j \geq j_m}})(|\alpha \cup \beta|)} & j < j_x \\
    1 & j \geq j_x.
    \end{cases}
\end{equation}

\paragraph{\cite[Proposition~\ref{IE-prop:1-LTdefXY}]{BS-rg-IE}}
As we explain next, using \refeq{cgamobs} and
identical notation to that defined in and around
\cite[Proposition~\ref{IE-prop:1-LTdefXY}]{BS-rg-IE},
the proposition holds as stated also for
the improved norms, provided we take
$A \ge 5+s$.
For this, what is required is to show that under the hypotheses of
\cite[Proposition~\ref{IE-prop:1-LTdefXY}]{BS-rg-IE},
the $\gamma_{\alpha,\beta}$ that arise in its proof obey
\begin{equation}
\lbeq{gamalphbet}
\gamma_{\alpha,\beta} \leq C
  \begin{cases}
  L^{-5} & |\alpha\cup\beta| = 0 \\
  L^{-1}  & |\alpha\cup\beta| = 1, 2.
  \end{cases}
\end{equation}
For $|\alpha\cup\beta| = 0$,
the first term of \refeq{cgamobs-L} obeys the bound of \refeq{gamalphbet},
since $d_\varnothing'=d+1$.
For the remaining cases, $d_\alpha'=2$ for $j < j_x$ and $d_\alpha'=1$ for $j \ge j_x$.
For $|\alpha\cup\beta| = 2$, the assumption that $F_1,F_2,F_1F_2$ have no component
in $\Ncal_{0x}$ unless $j \geq j_x$ means that we are in the case with no
growth due the ratio $\ell_{\sigma,j+1}/\ell_{\sigma,j}$ in \refeq{cgamobs-L},
and its first term again obeys the bound
\refeq{gamalphbet} with room to spare.
Finally, when $|\alpha\cup\beta| = 1$,
the first term of \refeq{cgamobs-L} also obeys the estimate
\refeq{gamalphbet}, and again with room to spare.
Concerning the second term of \refeq{cgamobs-L}, given
our choice of $A$ and the fact that we need only consider the growing factor in
\refeq{cgamobs-L} for $|\alpha\cup\beta|=1$, it suffices to observe that
\begin{equation}
L^{-(A + 1)}
L^{1 + s \1_{j \geq j_m}} \leq L^{- 5}.
\end{equation}
This completes the proof of the improved version of
\cite[Proposition~\ref{IE-prop:1-LTdefXY}]{BS-rg-IE}.

\begin{proof}[Proof of Proposition~\ref{prop:cl}]
We complete the proof of Proposition~\ref{prop:cl} by modifying the proof of
\cite[Proposition~\ref{IE-prop:cl}]{BS-rg-IE} above the mass scale.
The estimate \cite[\eqref{IE-e:Lkpfii-1zbis}]{BS-rg-IE}
follows from \cite[Proposition~\ref{IE-prop:ip}]{BS-rg-IE}
as an estimate in terms of the modified norm pair \eqref{e:npmass},
for which \cite[Proposition~\ref{IE-prop:ip}]{BS-rg-IE} was verified in Section~\ref{sec:stability2}.
The bound \cite[\eqref{IE-e:cl-1}]{BS-rg-IE} with improved $\gamma$ is obtained by applying the improved
version of \cite[Proposition~\ref{IE-prop:1-LTdefXY}]{BS-rg-IE}. In the remainder
of the proof of \cite[Proposition~\ref{IE-prop:cl}]{BS-rg-IE}, we specialise each
occurrence of $\Gcal$ to the case $\Gcal = G$ and we conclude by obtaining an
analogue of \cite[\eqref{IE-e:FXbdKzzz}]{BS-rg-IE} with $\tilde G$ replaced by
$G$ by applying Lemma~\ref{lem:mart} rather than \cite[Lemma~\ref{IE-lem:mart}]{BS-rg-IE}.

An additional detail is that it is required that we choose the parameter defining the space $\Ncal$ to
obey $p_\Ncal >A$.
Since we have changed $A$ (depending on $s$), we must make a corresponding change to $p_\Ncal$. This does not pose problems
(beyond the previously discussed requirement that $g$ needs to be chosen small depending on $p$),
as this parameter may be
fixed to be an arbitrary and sufficiently large integer
(see \cite[Section~\ref{phi4-sec:pNcal}]{ST-phi4} where this point is addressed
in a different context).  Similarly, the value of $A$ is immaterial and can be
any fixed number in the proof of
\cite[Proposition \ref{IE-prop:cl}]{BS-rg-IE}.
\end{proof}


\setcounter{section}{0}
\renewcommand{\thesection}{\Alph{section}}
\section{Moments of the free Green function}
\label{app:free-moments}

We now prove Proposition~\ref{prop:Gab-free-moment-estimate},
which we repeat
as the following proposition.

\begin{prop}\label{prop:Gab-free-moment-estimate-bis}
Let ${\sf c}_p$ be the constant defined by \eqref{e:cpdef}.
For all dimensions $d>2$ and all $p>0$,
as $m^2 \downarrow 0$,
\begin{equation}
\label{e:Gab-free-moment-estimate-bis}
\sum_{x\in\Zd} |x|^p G_{x}(0, m^2)
=
{\sf c}_p^p m^{-(p + 2)} (1 + O(m)).
\end{equation}
In particular, $\xi_p(0,\varepsilon) = {\sf c}_p \varepsilon^{-1/2}
(1+O(\varepsilon^{1/2}))$ as $\varepsilon \downarrow 0$.
\end{prop}

The last sentence in the the proposition follows immediately from
\refeq{Gab-free-moment-estimate-bis} and the fact that $\chi(0,m^2)=m^{-2}$,
so it suffices to prove \refeq{Gab-free-moment-estimate-bis}.

The case $p = 2$ of \refeq{Gab-free-moment-estimate-bis}
can be obtained easily from the identity
\begin{equation}
\sum_{x\in\Zd} |x|^2 G_x(0, m^2) = -\Delta_\Rd \hat{G}(0),
\end{equation}
where $\hat G$ is the Fourier transform of $G$.
Higher even moments could in principle
be computed by further differentiating $\hat G$.
We adopt a different approach
for general $p>0$,
based on the finite range decomposition of $(-\Delta_{\Zd}+m^2)^{-1}$
given in \cite{BGM04,Baue13a}.
This finite range decomposition also provides the basis for the renormalisation group method.
The finite range decomposition is
\begin{equation}
    G_x(0,m^2) = \sum_{j=1}^\infty C_{j;x}(m^2).
\end{equation}
The finite range property refers to the fact that $C_{j;x}(m^2) = 0$ if
$|x| \ge \frac 12 L^j$,
where $L>1$ is fixed arbitrarily.
 We review some properties of this decomposition, from
\cite{BBS-rg-pt,Baue13a}, before
proving Proposition~\ref{prop:Gab-free-moment-estimate-bis}.
The positive definiteness of the finite range
decomposition is not needed here, and $L$ need not be large.

The terms $C_{j;x}(m^2)$ are defined in \cite[Section~\ref{pt-sec:Cdecomp}]{BBS-rg-pt} by
\begin{equation} \label{e:Cdef}
  C_{j;x}(m^2) = \left\{\begin{aligned}
      &\int_{0}^{\frac{1}{2} L}  \phi_{t}^*(x; m^2) \; \frac{dt}{t}
      &\quad& (j=1)
      \\
      &\int_{\frac{1}{2} L^{j-1}}^{\frac{1}{2} L^{j}}  \phi_{t}^*(x; m^2) \; \frac{dt}{t}
      && (j\ge 2)
    \end{aligned}\right.
\end{equation}
(in \cite{BBS-rg-pt}, the notation $C_{j;0,x}$ and $\phi^*_t(0, x; m^2)$ was used instead).
Here, $\phi_t^*$ is a function of $x \in \Rd$ and $m^2 > 0$ given in
\cite[Example 1.1]{Baue13a}. It satisfies the finite range property that
$\phi_t^*(x; m^2) = 0$ for $|x| > t$.
It was also shown in \cite{Baue13a} that there exists a function $\phi_t$
satisfying the same finite range property but giving a decomposition of the
\emph{continuum} Green function:
\begin{align}
\label{e:frd-cont-int}
    (-\Delta_{\Rd} + m^2)^{-1}_{0x}
    &=
    \int_0^\infty \phi_t(x; m^2) \frac{dt}{t} .
\end{align}
Moreover, by \cite[(1.37)]{Baue13a}, for $|x| \leq t$,
\begin{equation}
\label{e:frd-Zd-Rd}
\phi^*_t(x; m^2) = \phi_t(x; m^2) + O(t^{-(d-1)} (1 + m^2 t^2)^{-k}).
\end{equation}
This allows us to approximate the discrete Green function by the continuum one, for which the moments are easily computed.
We have set the constant $c$ present in \cite{Baue13a} equal to $1$, which we can do by rescaling $\phi_t^*$.

As $t$ approaches $0$, the error bound in \eqref{e:frd-Zd-Rd} degenerates.
However, to estimate \eqref{prop:Gab-free-moment-estimate-bis}, it suffices to
restrict to $x \neq 0$.
Then, since $x \in \Z^d$, the finite range property permits replacement of the lower bound
in the range of integration for $j=1$ in \eqref{e:Cdef} by $\frac12$, and the contribution
due to $j=1$ can be estimated in the same way as the terms $j\geq 2$.

Also, by \cite[(1.34)]{Baue13a}, for any $k$ there is a constant $C_k$ such that
\begin{equation}
\label{e:frd-deriv-est}
|D_x \phi_t(x; m^2)| \leq C_k t^{-(d - 1)} (1 + m^2 t^2)^{-k}.
\end{equation}
We fix a choice of $k$ which obeys $k > \frac 12 (p+1)$ and use only this choice.
By \cite[(1.38)]{Baue13a}, there exists a function $\bar\phi$ such that
\begin{equation}
\label{e:frd-scaling}
\phi_t(x; m^2) = t^{-(d - 2)} \bar\phi\left(\frac{x}{t}; m^2 t^2\right).
\end{equation}

\begin{proof}[Proof of Proposition~\ref{prop:Gab-free-moment-estimate}]

We begin by writing
\begin{align}
\sum_{x\in\Zd} |x|^p G_x(0, m^2)
    &=
    \sum_{x\in\Zd} |x|^p
    \sum_{j=1}^\infty C_{j;x}(m^2)
    =
M(m^2)
+
E(m^2)
,
\end{align}
where the main and error terms are respectively
\begin{align}
M(m^2) &=
\sum_{x\in\Zd} |x|^p \sum_{j=1}^\infty
\int_{\frac 12 L^{j-1}}^{\frac 12 L^j} \phi_t(x; m^2) \;\frac{dt}{t},
\\
E(m^2) &= \sum_{x\in\Zd} |x|^p \sum_{j=1}^\infty \left(C_{j;x} - \int_{\frac{1}{2} L^{j-1}}^{\frac{1}{2} L^j} \phi_t(x; m^2) \frac{dt}{t}\right).
\end{align}

We first compute the main term $M$. By \eqref{e:frd-scaling},
\begin{equation}
\phi_t(x; m^2) = m^{d-2} \phi_{mt}(mx; 1).
\end{equation}
Therefore, by Riemann sum approximation,
\begin{align}
\sum_{x\in\Zd} &|x|^p \int_{\frac{1}{2} L^{j-1}}^{\frac{1}{2} L^j} \phi_t(x; m^2) \; \frac{dt}{t} \\
  &= m^{-(p+2)} m^d \sum_{x\in\Zd} |mx|^p \int_{\frac{1}{2} L^{j-1}}^{\frac{1}{2} L^j} \phi_{mt}(mx; 1) \; \frac{dt}{t}
  \\ \nonumber
  &= m^{-(p + 2)} \int_\Rd |x|^p \int_{\frac{1}{2} L^{j-1}}^{\frac{1}{2} L^j} \phi_{mt}(x; 1) \; \frac{dt}{t}
    + O(L^{(p + 1) j}
    L^{-2k(j-j_m)_+})
    ,
\end{align}
where the error estimate follows from \eqref{e:frd-deriv-est} and \eqref{e:mass-decay}.
Summation over $j$ gives
\begin{align}
M(m^2)
&= {\sf c}_p^p m^{-(p + 2)} + O(m^{-(p + 1)}),
\end{align}
where we used
\eqref{e:frd-cont-int} for the first term, and we used $2k>p+1$ and
Lemma~\ref{lem:mass-scale-sum} for the second term.

For the error term,
it follows from
\eqref{e:Cdef}, \eqref{e:frd-Zd-Rd},
and the observation that the lower bound in the range of integration for the $j=1$ term in \eqref{e:Cdef}
can be changed to $\frac12$ that
\begin{equation}
\label{e:frd-integral-approx}
C_{j;x}
=
\int_{\frac{1}{2} L^{j-1}}^{\frac{1}{2} L^j} \phi_t(x; m^2) \frac{dt}{t}
+
O(L^{-j (d - 1)} (1 + m^2 L^{2j})^{-k})
\1_{|x|\leq L^j}.
\end{equation}
Therefore, again using \eqref{e:mass-decay}, we have
\begin{align}
E(m^2)
&= \sum_{j=1}^\infty \sum_{|x|\leq L^j} |x|^p O(L^{-j (d - 1)}
L^{-2k(j-j_m)_+})
\\
&= \sum_{j=1}^\infty O(L^{(p+1)j}L^{-2k(j-j_m)_+})
\label{e:free-error-estimate}
.
\end{align}
With $2k>p+1$ and Lemma~\ref{lem:mass-scale-sum}, this gives
$E(m^2) = O(m^{-(p+1)})$,
and the proof is complete.
\end{proof}


\section*{Acknowledgements}
The work of GS, AT, BW was supported in part by NSERC of Canada.
The authors are grateful to David Brydges for important conversations and advice,
and to an anonymous referee for useful suggestions.

\bibliography{rg}

\begin{thebibliography}{10}

\bibitem{Baue13a}
R.~Bauerschmidt.
\newblock A simple method for finite range decomposition of quadratic forms and
  {Gaussian} fields.
\newblock {\em Probab. Theory Related Fields}, {\bf 157}:817--845, (2013).

\bibitem{BBS-phi4-log}
R.~Bauerschmidt, D.C. Brydges, and G.~Slade.
\newblock Scaling limits and critical behaviour of the $4$-dimensional
  $n$-component $|\varphi|^4$ spin model.
\newblock {\em J. Stat. Phys}, {\bf 157}:692--742, (2014).

\bibitem{BBS-saw4}
R.~Bauerschmidt, D.C. Brydges, and G.~Slade.
\newblock Critical two-point function of the 4-dimensional weakly self-avoiding
  walk.
\newblock {\em Commun.\ Math.\ Phys.}, {\bf 338}:169--193, (2015).

\bibitem{BBS-saw4-log}
R.~Bauerschmidt, D.C. Brydges, and G.~Slade.
\newblock Logarithmic correction for the susceptibility of the 4-dimensional
  weakly self-avoiding walk: a renormalisation group analysis.
\newblock {\em Commun.\ Math.\ Phys.}, {\bf 337}:817--877, (2015).

\bibitem{BBS-rg-pt}
R.~Bauerschmidt, D.C. Brydges, and G.~Slade.
\newblock A renormalisation group method. {III}. {Perturbative} analysis.
\newblock {\em J. Stat. Phys}, {\bf 159}:492--529, (2015).

\bibitem{BBS-rg-flow}
R.~Bauerschmidt, D.C. Brydges, and G.~Slade.
\newblock Structural stability of a dynamical system near a non-hyperbolic
  fixed point.
\newblock {\em Ann. Henri Poincar\'e}, {\bf 16}:1033--1065, (2015).

\bibitem{BGZ73}
E.~Br\'ezin, J.C. Le~Guillou, and J.~Zinn-Justin.
\newblock Approach to scaling in renormalized perturbation theory.
\newblock {\em Phys.\ Rev.\ D}, {\bf 8}:2418--2430, (1973).

\bibitem{BGM04}
D.C. Brydges, G.~Guadagni, and P.K. Mitter.
\newblock Finite range decomposition of {Gaussian} processes.
\newblock {\em J. Stat. Phys.}, {\bf 115}:415--449, (2004).

\bibitem{BI03c}
D.C. Brydges and J.Z. Imbrie.
\newblock End-to-end distance from the {G}reen's function for a hierarchical
  self-avoiding walk in four dimensions.
\newblock {\em Commun. Math. Phys.}, {\bf 239}:523--547, (2003).

\bibitem{BS-rg-norm}
D.C. Brydges and G.~Slade.
\newblock A renormalisation group method. {I}. {Gaussian} integration and
  normed algebras.
\newblock {\em J. Stat. Phys}, {\bf 159}:421--460, (2015).

\bibitem{BS-rg-loc}
D.C. Brydges and G.~Slade.
\newblock A renormalisation group method. {II}. {Approximation by local
  polynomials}.
\newblock {\em J. Stat. Phys}, {\bf 159}:461--491, (2015).

\bibitem{BS-rg-IE}
D.C. Brydges and G.~Slade.
\newblock A renormalisation group method. {IV}. {Stability} analysis.
\newblock {\em J. Stat. Phys}, {\bf 159}:530--588, (2015).

\bibitem{BS-rg-step}
D.C. Brydges and G.~Slade.
\newblock A renormalisation group method. {V}. {A} single renormalisation group
  step.
\newblock {\em J. Stat. Phys}, {\bf 159}:589--667, (2015).

\bibitem{FFS92}
R.~Fern\'{a}ndez, J.~Fr\"{o}hlich, and A.D. Sokal.
\newblock {\em Random Walks, Critical Phenomena, and Triviality in Quantum
  Field Theory}.
\newblock Springer, Berlin, (1992).

\bibitem{Hara87}
T.~Hara.
\newblock A rigorous control of logarithmic corrections in four dimensional
  $\varphi^4$ spin systems. {I}. {Trajectory} of effective {Hamiltonians}.
\newblock {\em J. Stat. Phys.}, {\bf 47}:57--98, (1987).

\bibitem{HT87}
T.~Hara and H.~Tasaki.
\newblock A rigorous control of logarithmic corrections in four dimensional
  $\varphi^4$ spin systems. {II}. {Critical} behaviour of susceptibility and
  correlation length.
\newblock {\em J. Stat. Phys.}, {\bf 47}:99--121, (1987).

\bibitem{LK69}
A.I. Larkin and D.E. Khmel'Nitski\u{i}.
\newblock Phase transition in uniaxial ferroelectrics.
\newblock {\em Soviet Physics JETP}, {\bf 29}:1123--1128, (1969).
\newblock {English} translation of {\it Zh.\ Eksp.\ Teor.\ Fiz.} {\bf 56},
  2087--2098, (1969).

\bibitem{ST-phi4}
G.~Slade and A.~Tomberg.
\newblock Critical correlation functions for the $4$-dimensional weakly
  self-avoiding walk and $n$-component $|\varphi|^4$ model.
\newblock {\em Commun. Math. Phys.}, {\bf 342}:675--737, (2016).

\bibitem{WR73}
F.J. Wegner and E.K. Riedel.
\newblock Logarithmic corrections to the molecular-field behavior of critical
  and tricritical systems.
\newblock {\em Phys. Rev. B}, {\bf 7}:248--256, (1973).

\end{thebibliography}
\bibliographystyle{plain}

\end{document}